\documentclass[10pt,a4paper]{article}

\usepackage{setspace}
\usepackage{amsmath, latexsym, amssymb}
\numberwithin{equation}{section}
\usepackage[english]{babel}
\usepackage{color}
\definecolor{myred}{rgb}{0.9,0,0}
\definecolor{myblue}{rgb}{0,0,0.8}
\definecolor{mygreen}{rgb}{0,0.8,0}
\definecolor{gray}{rgb}{0.9,0.9,0.9}
\usepackage{ifpdf}

\ifpdf
   \usepackage{graphicx}
    \usepackage[pdftex,
               pdftitle={Analytical Framework for Credit Portfolios},
               pdfsubject={Risk measurement and management},
               pdfkeywords={analytical VaR, credit portfolio, capital allocation, multi-factor model},
               pdfauthor={
               Mikhail Voropaev
               },
               pdfstartview=FitH,
               bookmarks=false,
               breaklinks=true,
               colorlinks=true,
               citecolor=mygreen,
               linkcolor=myred,
               urlcolor=myblue
               ]{hyperref}
\else
   \usepackage{hyperref}
   \usepackage[dvips]{graphicx}
   \usepackage{rotating}
\fi

\usepackage{listings}
\usepackage[authoryear, round, sort]{natbib}

\begin{document}
\title{Analytical Framework for Credit Portfolios}
\author{Mikhail Voropaev
\thanks{Independent consultant. E-mail:
\href{mailto:mvoropaev@mail.ru}{mvoropaev@mail.ru}.
}
}
\date{July 2010}
\maketitle

\begin{abstract}
Analytical, free of time consuming Monte Carlo simulations, framework for credit portfolio systematic risk metrics calculations is presented. Techniques are described that allow calculation of portfolio-level systematic risk measures (standard deviation, VaR and Expected Shortfall) as well as allocation of risk down to individual transactions. The underlying model is the industry standard multi-factor Merton-type model with arbitrary valuation function at horizon (in contrast to the simplistic default-only case). High accuracy of the proposed analytical technique is demonstrated by benchmarking against Monte Carlo simulations.
\end{abstract}

\section{Introduction}
There exists an increasing demand for fast and consistent economic capital calculation and allocation techniques. Portfolio-wide calculations of economic capital are just a first step in the modern process of credit portfolio management. Financial institutions are more and more involved in stress testing, sensitivity and scenario analysis. For these purposes the portfolio-level risk measures need to be recalculated over and over again. Using industry standard Monte Carlo simulations for the portfolio-level risk quantification requires considerable amount of time and computer power. For the purposes of risk concentration identification, risk-adjusted pricing and portfolio optimization the portfolio-wide risk (economic capital) needs to be allocated down to individual transactions. The latter task is even more challenging from both methodological and computational points of view. Statistical noise, being inherent part of Monte Carlo simulations, leads to unstable estimations of the allocated risk (especially in case of VaR-based capital allocation). Reliable estimations of capital charges based on simulations require significantly more computer time/power compared to the portfolio-wide calculations.

Although several techniques have been developed to improve the performance of the simulations-based approach, e.g. importance sampling \citep[see, e.g.,][]{Kalkbrener2004} and kernel estimators \citep[see, e.g.,][]{KernelEstimators}, simulation-based estimation of risk contributions on transaction level is still a demanding computational problem. In practice, when applied to multi-factor models, efficiency of the abovementioned techniques may be limited since it depends on a quality of analytical approximation used to determine a sampling region. Yet another drawback of the simulation-based approach is its inability to efficiently risk-assess new deals in a context of the portfolio.

An alternative to the simulation-based approach would be some kind of analytical technique. Although Merton-type models are not analytically tractable in general case, some progress has been made to develop an approximate solution. The most successful attempts to tackle the problem are \emph{Asymptotic Single Risk Factor} (ASRF) framework \citep[][]{ASRF}, \emph{granularity adjustment} (GA) by \citet{Unsystematic} and \citeauthor{MultiFactor}'s \citeyearpar{MultiFactor} \emph{multi-factor adjustment}. This article aims to complement the existing analytical techniques. The ambition is to fill the existing gap between theoretical results and practice by considering a fully-featured PortfolioManager-type \citep[][]{PortfolioManager} credit portfolio model. The proposed framework allows to calculate most commonly used risk measures (variance, value-at-risk and expected shortfall) on both portfolio and transaction levels. The material presented here lacks mathematical rigor. Instead, results of numerical tests are presented to demonstrate the performance and prove the validity of the proposed techniques.

This article is organized as follows. First, in Section \ref{sec:structural}, a short description of the multi-factor Merton-type model is given, followed by a review of the progress made so far on the model's analytical tractability. VaR expansion technique, used as a starting point for the approach presented here, is presented in Section \ref{sec:adjustments}. Main results are presented in Section \ref{sec:systematic}, where conditional expectation series expansion is derived (Sections \ref{sec:sf} and \ref{sec:mf}) and applied to systematic risk constituents. It is demonstrated, how the proposed expansion technique can be utilized to compute systematic components of various portfolio-wide risk measures and corresponding risk contributions. Finally, in Section \ref{sec:idiosyncratic}, the conditional expectation series expansion technique is extended to cover idiosyncratic risk components. Monte Carlo simulations are used to substantiate the validity of the proposed analytical approach (Sections \ref{sec:numerical} and \ref{sec:numerical2}).

\section{Structural credit portfolio models}\label{sec:structural}
Merton-type credit portfolio models are most widely accepted ones for the purposes of credit portfolio risk metrics calculations. In these models the portfolio consists of risky instruments $\{v_i\}$ with the
value $v_i$ of each instrument at horizon (usually set to one year) being a function of normally distributed random variable $\epsilon_i$ (normalized asset return). Correlations between  these variables $\{\epsilon_i\}$ are modeled through a set of $N_f$ normally distributed independent variables $\{\eta_k\}$ referred to as common factors.
Each variable $\epsilon_i$ is split in a sum of instrument specific (idiosyncratic) part, which depends on a Gaussian variable $\xi_i$, and systematic part, which depends on the common factors, as follows
\begin{equation}\label{eq:Merton}
v_i(\epsilon_i) = v_i\bigl(\rho_i{\textstyle\sum_k}(\beta_i)_k\eta_k + {\textstyle\sqrt{1-\rho_i^2}~\xi_i}\bigr).
\end{equation}
The independently distributed random variables $\{\{\xi_i\},\{\eta_k\}\}$\footnote{Assuming $\{\xi_i\}$ to be independently distributed is equivalent to an assumption that each borrower in the portfolio is represented by one facility. This assumption is made to simplify notations and does not undermine the validity of the results.} are assumed to have zero mean and unit variance. Instrument specific constants $|\rho_i|<1$ and $\{(\beta_i)_k\}$ determine dependency of $\epsilon_i$ on the common factors (related to geographic regions and industry types).
The so-called factor loadings $\{(\beta_i)_k\}$ are subject to normalization condition
\begin{equation}\label{eq:betanorm}
{\textstyle\sum_k}(\beta_i)_k^2 = 1.
\end{equation}

Uncertainty in the value of the portfolio $V = \sum_iv_i$ is quantified by means of various \emph{risk measures}, most popular of which are VaR(Value-at-Risk), ES(Expected Shortfall) and standard deviation\footnote{See \citet{RiskMeasures} for a detailed discussion of the various risk measures.}.

Once the portfolio-level risk measure is known, the question arises how to distribute (allocate) this risk consistently among the constituents. The Euler allocation technique \cite[see, e.g.,][]{Euler} is the commonly adopted solution. According to the Euler allocation principle, individual assets $v_i$ of the portfolio are assigned fractions (risk contributions) $\theta_i$ of the portfolio-level risk $\Theta$ according to
\begin{equation}\label{eq:Euler}
\theta_i = w_i\frac{\partial\Theta}{\partial w_i}, \qquad \Theta = \sum_i\theta_i,
\end{equation}
where $w_i$ is a weight of $i$th facility in the portfolio. In what follows the weights $\{w_i\}$ will be implied but not written explicitly.

No closed-form solution exists for either portfolio-level or facility-level risk measures in the general case. Several important steps have been made towards approximate analytical solution of the problem. First, the case of one common factor and infinitely large and fine-grained portfolio was solved by Asymptotic Single Risk Factor framework \citep[][]{ASRF}. Next, idiosyncratic component of risk has been addressed by granularity adjustment \citep[][]{Unsystematic}. Finally, the results of \citet{Unsystematic} were applied to a multi-factor case by \citet{MultiFactor}.

Unfortunately, no significant progress has been made ever since towards better analytical approximation; however, some generalizations of previous results have been recently reported by \citet{GAmtm} and \citet{GAmf} and attempts have been made to find a more simple solution to the multi-factor case by \citet{Duellmann} and \citet{Cespedes}. Moreover, practitioners considering applying Pykhtin's approach to realistic credit portfolio models face two major difficulties. First, Pykhtin's model was formulated for a default-only case and it is not at all obvious how to (efficiently) extend it to a more general (and realistic) case of value-based valuation at horizon. Second, calculation of the multi-factor adjustment are of quadratic in portfolio size complexity, making application of the model to large portfolios barely possible. On top of that, no solution to the problem of risk allocation within Pykhtin's model has ever been reported.

In the following sections a new approach is presented. Although based on the same principles, the approach will address the above mentioned difficulties of Pykhtin's model. Moreover, higher order (i.e. third order vs. original second order) multi-factor adjustments will be considered. Many of the restrictive simplifications (e.g. default-only mode, simplistic correlation structure, homogeneous portfolio, etc.) commonly used in previous publications will be loosened, allowing application of the proposed analytical framework to a wide class of realistic\footnote{"realistic" here means "used in practice".} structural credit portfolio models.

\section{VaR and ES adjustments}\label{sec:adjustments}
Building on the work of \citet{VaRderivatives}, \citet{Unsystematic} derived the second order correction to VaR and used the results in the context of credit portfolio to calculate an adjustment for undiversified idiosyncratic risk (\emph{granularity adjustment}). Somewhat simpler derivation is presented here, outcome of which is a higher precision correction to VaR and is more suitable for the techniques presented in this article.

Consider random variable $x$ with continuous probability distribution function (p.d.f.) $f(x)$. Let $q_\alpha$ be the $\alpha$-quantile of this distribution. Consider another random variable $\delta x$ with $g(\delta x|x)$ being its p.d.f. conditional on the value of the first variable $x$. Let us find the $\alpha$-quantile $q^*_\alpha$ of the p.d.f. $f^*(x+\delta x)$ of the sum of the above two variables. The $f^*$ can be written as
\begin{equation}
f^*(x) = \int f(x-\delta x)g(\delta x|x-\delta x)\mathrm{d}(\delta x)
\end{equation}
Expanding the right hand side of this expression in Taylor series of $(x-\delta x)$ around $x$, one can obtain
\begin{equation}\label{eq:pdfexp}
f^*(x) = f(x) + \sum_{n=1}^\infty \frac{(-1)^n}{n!}\frac{\mathrm{d}^n}{\mathrm{d}x}[f(x)\mu_n(x)], \quad \mu_n(x) = \int (\delta x)^ng(\delta x|x)\mathrm{d}(\delta x),
\end{equation}
where $\mu_n(x)$ are moments of $\delta x$ distribution conditional on $x$.

Once the relationship \eqref{eq:pdfexp} between probability distribution functions has been established, the relationship between quantiles can be derived by substituting \eqref{eq:pdfexp} into the following definition of $\alpha$-quantile
\begin{equation}
\alpha = \int_{-\infty}^{q_\alpha} f(x)\mathrm{d}x = \int_{-\infty}^{q^*_\alpha} f^*(x)\mathrm{d}x
\end{equation}
The result is
\begin{equation}\label{eq:varexp}
\int^{q^*_\alpha}_{q_\alpha}f(x)\mathrm{d}x = \sum_{n=1}^\infty\frac{(-1)^{n-1}}{n!}\frac{\mathrm{d}^{n-1}}{\mathrm{d}x^{n-1}}[f(x)\mu_n(x)]\Big|_{x=q_\alpha^*}
\end{equation}
Suppose $\delta x$ is a small correction to $x$. One way to quantify this smallness is to
assume that $\mu_n \sim \delta^n$, where $\delta$ is some small number. One can solve the equation \eqref{eq:varexp} order by order in $\delta$ by expanding its both sides in powers of
$(q^*_\alpha - q_\alpha)$ around $q_\alpha$.

Only distributions satisfying $\mu_1(x)\equiv 0$ will be considered in this article. In this case the $\{\mu_n(x)\}$ become conditional \emph{central} moments and \eqref{eq:varexp} has a particularly simple third order solution
\begin{equation}\label{eq:varadj}
q^*_\alpha - q_\alpha \approx -\frac{1}{2f(x)}\frac{\mathrm{d}}{\mathrm{d}x}[f(x)\mu_2(x)]\Big|_{x=q_\alpha}
+ \frac{1}{6f(x)}\frac{\mathrm{d}^2}{\mathrm{d}x^2}[f(x)\mu_3(x)]\Big|_{x=q_\alpha}
\end{equation}

Let us look at the result \eqref{eq:varadj} from credit portfolio perspective. Let $x$ be a single factor approximation of the portfolio value, $x=V(\eta)$. Let the factor $\eta$ be normally distributed with the p.d.f. $n(\eta) = e^{-\eta^2/2}/\sqrt{2\pi}$. The $\alpha$-quantile $q_\alpha$ is related to the portfolio's VaR and portfolio's expected value $\text{E}(V)$ as\footnote{VaR defined this way is simply an \emph{economic capital} of the portfolio.}
\begin{equation}\label{VaRvsquantile}
\text{VaR} = \text{E}(V) - q_\alpha
\end{equation}

Using $n'(\eta) = -\eta n(\eta)$, $f(V)\text{d}V = n(\eta)\mathrm{d}\eta$ and \eqref{eq:varadj}, the second and third order VaR adjustments can be written as\footnote{The signs of both VaR and ES adjustments seem different from those that can be found in the literature. This apparent contradiction is explained by the fact that the analysis here is based on the value of the portfolio $V$, rather than its losses.}
\begin{eqnarray}
\Delta \text{VaR}_2(\alpha) & = & \frac{1}{2\,n(\eta)} \frac{\text{d}}{\text{d}\eta} \left(\frac{n(\eta)\mu_2(\eta)}{V'(\eta)} \right)\Big|_{\eta=\Phi^{-1}(\alpha)}  = \label{eq:VaR2} \\
& = & \frac{1}{2V'}\left( \mu_2' - \mu_2\left(\eta + \frac{V''}{V'}\right)\right)\Big|_{\eta=\Phi^{-1}(\alpha)} \nonumber \\
\Delta \text{VaR}_3(\alpha) & = & -\frac{1}{6\,n(\eta)} \frac{\text{d}}{\text{d}\eta} \left(\frac{1}{V'(\eta)}\frac{\text{d}}{\text{d}\eta}\left(\frac{n(\eta)\mu_3(\eta)}{V'(\eta)} \right)\right)\Big|_{\eta=\Phi^{-1}(\alpha)}  = \label{eq:VaR3} \\
& = & -\frac{1}{6[V']^2}\left(\mu_3'' - \mu_3'\left(2\eta + 3\frac{V''}{V'}\right) +
\mu_3\left((\eta^2-1) + 3\eta\frac{V''}{V'} + \frac{3[V'']^2  - V'V'''}{[V']^2}\right)\right)\Big|_{\eta=\Phi^{-1}(\alpha)}\nonumber
\end{eqnarray}
where $\Phi^{-1}(\alpha)$ is the inverse of the normal cumulative p.d.f.

Using the VaR adjustments \eqref{eq:VaR2} and \eqref{eq:VaR2}, one can easily calculate similar adjustments to expected shortfall. Noticing that
\begin{equation}
\text{ES}(\alpha) = \frac{1}{\alpha}\int_{-\infty}^{\eta=\Phi^{-1}(\alpha)}\text{VaR}(\eta)n(\eta)\mathrm{d}\eta,
\end{equation}
the second and third order expected shortfall contributions can be written as
\begin{eqnarray}
\Delta \text{ES}_2(\alpha) & = & \frac{1}{2\alpha}\frac{n}{V'}\mu_2\Big|_{\eta=\Phi^{-1}(\alpha)}  \label{eq:ES2} \\
\Delta \text{ES}_3(\alpha) & = & -\frac{1}{6\alpha} \frac{1}{V'}\frac{\text{d}}{\text{d}\eta}\left(\frac{n\mu_3}{V'} \right)\Big|_{\eta=\Phi^{-1}(\alpha)}  = \label{eq:ES3} \\
& = & -\frac{1}{6\alpha}\frac{n}{[V']^2}\left(\mu_3' - \mu_3\left(\eta + \frac{V''}{V'} \right)\right)\Big|_{\eta=\Phi^{-1}(\alpha)} \nonumber
\end{eqnarray}

\section{Systematic risk}\label{sec:systematic}
Let us start by assuming that the portfolio dynamics is mainly governed by the systematic risk components, i.e. the common factors $\{\eta_k\}$ give the main contribution to the portfolio risk measures, while the idiosyncratic factors $\{\xi_i\}$ give rise to less significant corrections. In this section it will be demonstrated, how to isolate the systematic risk by integrating out the idiosyncratic components. It will also be shown, how the resulting asymptotic multi-factor framework can be utilized for the risk metrics calculations.

The same framework will be further extended in the next section to cover the idiosyncratic risk components.

\subsection{Series expansion for conditional expectation: single factor}\label{sec:sf}
In order to focus on the systematic part of portfolio dynamics, let us integrate out (average over) the idiosyncratic component $\xi_i$ in \eqref{eq:Merton}. Let us assume there is just one common factor and extend the results to a multi-factor case later.

Average value of a facility $\overline{v_i}$ conditional on the systematic factor $\eta$ is
\begin{equation}
\overline{v_i}(\eta) = \int\! v_i(\rho_i\eta + {\textstyle\sqrt{1-\rho_i^2}}\,\xi)\frac{e^{-\xi^2/2}}{\sqrt{2\pi}}\mathrm{d}\xi,
\end{equation}
which after changing the integration variable to $\epsilon = \rho_i\eta + \sqrt{1-\rho_i^2}\,\xi$ becomes
\begin{equation}\label{eq:notawowyet}
\overline{v_i}(\eta) = \int\!v(\epsilon)\,\frac{1}{\sqrt{1-\rho_i^2}}\,\mathrm{exp}\left(\frac{2\rho_i\epsilon\eta - \rho_i^2(\epsilon^2+\eta^2)}{2(1-\rho_i^2)}\right)\frac{e^{-\epsilon^2/2}}{\sqrt{2\pi}}\text{d}\epsilon.
\end{equation}
The above expression can be further developed by applying Mehler's formula \citep[for the proof see, e.g.,][]{Mehler}:
\begin{equation}
\sum_{n=0}^\infty\mathrm{He}_n(\epsilon)\mathrm{He}_n(\eta)\frac{\rho^n}{n!} = \frac{1}{\sqrt{1-\rho^2}} \mathrm{exp}\left(\frac{2\rho\epsilon\eta - \rho^2(\epsilon^2+\eta^2)}{2(1-\rho^2)}\right),
\end{equation}
where $\text{He}_n(\eta) = (-1)^ne^{\eta^2/2}(d/d\eta)^ne^{-\eta^2/2}$ are Hermite polynomials \citep[for definition and properties of Hermite polinomials see, e.g.,][]{Abramowitz}. The result is
\begin{equation}\label{eq:wow}
\overline{v_i}(\eta) = \sum_n\frac{\rho_i^n}{n!}v_i^{(n)}\text{He}_n(\eta), \qquad v_i^{(n)} = \int\!v_i(\epsilon)\text{He}_n(\epsilon)\frac{e^{-\epsilon^2/2}}{\sqrt{2\pi}}\text{d}\epsilon
\end{equation}

A few remarks are needed regarding the result \eqref{eq:wow}. First, expansion exists as long as all the coefficients $v^{(n)}$ are finite. This is the case, for example, for any piece-wise continuous function $v_i(\epsilon)$ whose absolute value at infinity ($\epsilon \to \pm\infty$) does not increase faster than some power of $\epsilon$. Any reasonable value function of a financial instrument does satisfy this constrain.

Next, in case $\rho=1$, the classical Hermite series expansion is recovered. The series converges to the value of the function everywhere except for discontinuity points where the series converges to the average of the function around the discontinuity point. The Hermite series expansion is known for its slow convergence especially for large values of the argument $\eta$.

Finally, as a consequence of $|\rho|<1$ in \eqref{eq:wow}, the conditional expectation series converge significantly better. For the same reason, i.e. $|\rho|<1$, the conditional expectation function $\overline{v_i}(\eta)$ is not only continuous, but differentiable infinite number of times.

Before generalizing the result \eqref{eq:wow} to a multi-factor case, let us explore the benefits of the expansion \eqref{eq:wow} in the context of credit portfolio. Advantages of the proposed approach can be seen even in a simple case of a single factor model.

 The asymptotic single risk factor $\eta$ value of the portfolio $V_{1f}(\eta)=\sum_i\overline{v_i}(\eta)$ can be easily derived from \eqref{eq:wow} and is
\begin{equation}\label{eq:port1f}
V_{1f}(\eta) = \sum_nV^{(n)}\text{He}_n(\eta), \qquad V^{(n)} = \sum_i\frac{\rho_i^n}{n!}v_i^{(n)}
\end{equation}
Once the coefficients $V^{(n)}$ are calculated, one can immediately write both VaR and ES of the portfolio for any confidence level $\alpha$ as
\begin{equation}\label{eq:1fvares}
\text{VaR}(\alpha) = -\sum_{n>0}V^{(n)}\text{He}_n(\eta)\Big|_{\eta=\Phi^{-1}(\alpha)} \qquad
\text{ES}(\alpha) = \frac{e^{-\eta^2/2}}{\sqrt{2\pi}}\sum_{n>0}V^{(n)}\text{He}_{n-1}(\eta)\Big|_{\eta=\Phi^{-1}(\alpha)}
\end{equation}
Using \eqref{eq:Euler} and \eqref{eq:port1f}, trivial calculations lead to the following VaR and ES -based risk contributions
\begin{equation}\label{eq:1fvaresalloc}
\text{VaR}_i^c = -\sum_{n>0}\frac{\rho_i^n}{n!}v^{(n)}\text{He}_n(\eta)\Big|_{\eta=\Phi^{-1}(\alpha)} \qquad
\text{ES}_i^c(\alpha) = \frac{e^{-\eta^2/2}}{\sqrt{2\pi}}\sum_{n>0}\frac{\rho_i^n}{n!}v^{(n)}\text{He}_{n-1}(\eta)\Big|_{\eta=\Phi^{-1}(\alpha)}
\end{equation}

\subsection{Series expansion for conditional expectation: multiple factors}\label{sec:mf}
In a multi-factor case, the conditional expectation \eqref{eq:wow} can be written as
\begin{equation}
\overline{v_i}(\eta_k) = \sum_n\frac{\rho_i^n}{n!}v_i^{(n)}\text{He}_n\big({\textstyle\sum_k}(\beta_i)_k\eta_k\big)
\end{equation}
This expression, however, does not allow to write the portfolio value $V$ in a form similar to \eqref{eq:port1f}. To accomplish this, let us introduce multivariate Hermite polynomials
\begin{equation}\label{eq:multHermite}
\text{He}_n^{\overbrace{k_1k_2\ldots}^n}(\eta_k) = (-1)^n\exp{\Big(\frac{1}{2}\sum_m\eta_m^2\Big)} \overbrace{\frac{\partial}{\partial\eta_{k1}}\frac{\partial}{\partial\eta_{k2}}\ldots}^n\, \exp{\Big(-\frac{1}{2}\sum_m\eta_m^2\Big)}
\end{equation}
The multi-factor expansion then becomes
\begin{equation}\label{eq:wowmult}
\overline{v_i}(\eta_k) = \sum_n\frac{\rho_i^n}{n!}v_i^{(n)}\overbrace{(\beta_i)_{k_1}(\beta_i)_{k_2}\ldots}^n
\text{He}_n^{\overbrace{k_1k_2\ldots}^n}(\eta_k)
\end{equation}
and the conditional expectation of the portfolio can be written as
\begin{equation}\label{eq:portmult}
V(\eta_k) = \sum_n\sum_{\underbrace{k1,k2,\ldots}_n}V^{(n)}_{\underbrace{k_1k_2\ldots}_n}\text{He}^{\overbrace{k_1k_2\ldots}^n}_n(\eta), \qquad V_{\underbrace{k_1k_2\ldots}_n}^{(n)} = \sum_i\frac{\rho_i^n}{n!}v_i^{(n)}\overbrace{(\beta_i)_{k_1}(\beta_i)_{k_2}\ldots}^n
\end{equation}

Using orthogonality properties of multivariate Hermite polynomials
\begin{equation}\label{eq:hermiteorth}
\int\text{He}^{k_1k_2\ldots}_n(\eta_k)\text{He}^{l_1l_2\ldots}_m(\eta_k)\frac{e^{-\sum_{k=1}^{N_f}\eta_k^2/2}}{(2\pi)^{N_f/2}}
\mathrm{d}\eta_k = n!\,\delta_{nm}\delta_{k_1l_1}\delta_{k_2l_2}\ldots,
\end{equation}
one can calculate the variance $\sigma_V^2$ of the portfolio
\begin{equation}\label{eq:variance}
\sigma_V^2 = \text{E}(V^2) - (\text{E}(V))^2= \sum_{n>0}n!\sum_{k_1,k_2,\ldots}\Big[V^{(n)}_{k_1,k_2,\ldots}\Big]^2
\end{equation}
Standard deviation $\sigma_V$ based risk contributions can be calculated using the \eqref{eq:Euler} and \eqref{eq:portmult}. The result is
\begin{equation}\label{eq:varalloc}
\sigma_i^c = \frac{1}{\sigma_V}\sum_{n>0}\rho_i^nv_i^{(n)} \sum_{k_1,k_2,\ldots}(\beta_i)_{k_1}(\beta_i)_{k_2}\ldots V^{(n)}_{k_1,k_2,\ldots}
\end{equation}

Recently, it was shown by \citet{Voropaev} that applying \eqref{eq:variance} and \eqref{eq:varalloc} results in calculations which are of linear complexity in portfolio size. The amount of common factors $N_f$, however, is the bottleneck of the calculations. Indeed, $n$th term in the above expressions contains $N_f^n$ elements, making calculations of higher order terms impractical. Fortunately, only a few first terms lead to an accurate results. For details and discussion of the convergence properties of \eqref{eq:variance} the reader is referred to \citet{Voropaev}, where the problem of standard deviation and standard deviation base risk allocation has been solved in more general case using techniques similar to those described here. From now on we will focus on the tail risk measures, VaR and ES.

\subsection{Conditional expectation in the tail}\label{sec:systail}
Let us assume that the portfolio value distribution in the multi-factor case can be approximated by some single-factor value distribution, i.e. let us write the value of the portfolio as
\begin{equation}\label{eq:1f+mf}
V = V_{1f}(\vec{Y}) + V_{mf}, \qquad \text{E}(V_{mf}|V_{1f}) = 0,
\end{equation}
where $V_{1f}$ is a single-factor approximation and $V_{mf}$ is a multi-factor correction with zero expectation conditional on $V_{1f}$. The single systematic risk factor $\vec{Y}$ is a linear combination of the common factors $\{\eta_k\}$. The choice of the \emph{principal risk factor} $\vec{Y}$ is somehow arbitrary; however, one would aim to choose $\vec{Y}$ such that $V_{1f}$ is as good approximation to $V$ as possible and $V_{mf}$ is as small correction as possible. A solution to this optimization problem (which needs to be well formulated first) may be a matter of future research. Fortunately, as we will see later, even in case of sub-optimal choice of $\vec{Y}$ one can achieve very good numerical results.

The (sub-optimal) choice of $\vec{Y}$ used here is based on the following rationale. Notice that $n$th term in the conditional expectation expansion \eqref{eq:portmult} is (roughly speaking) proportional to $\rho_p^n$, where $\rho_p$ is some characteristic correlation. Assuming $\rho_p$ is small, one can conclude that the lower order terms in \eqref{eq:portmult} give the main contribution to the portfolio dynamics. Assuming further that the $n=1$ term is the most important one, one would naturally choose $\vec{Y}$ to point in the direction defined by
\begin{equation}
\vec{V}^{(1)} = (V^{(1)}_1, V^{(1)}_2, \ldots, V^{(1)}_{N_f})
\end{equation}
This particular choice of $\vec{Y}$ is not only natural and convenient within the proposed framework, but also will be substantiated by numerical tests in Section \ref{sec:numerical}\footnote{\citet{MultiFactor} suggests different choices for $\vec{Y}$. These choices, however, are intuition-based and are not theoretically substantiated. In the author's experience, the choice of $\vec{Y}$ presented here leads to better results when applied to realistic portfolios.}.

One last preparation is needed before splitting the portfolio value according to \eqref{eq:1f+mf}. Once the principal risk factor $\vec{Y}$ is known, let us transform the initial orthonormal set of common factors $\{\eta_k\}$ by some orthogonal transformation in such way, that one of the transformed factors coincides with $\vec{Y}$. This can be achieved by Gram-Schmidt process starting with $\vec{Y}$. From now on we will assume that the transformation took place and that $\{\eta_k\}$ is a set of the transformed common factors. The $\eta_1$ factor is assumed to be the principal risk factor.

To split the portfolio value \eqref{eq:portmult} according to \eqref{eq:1f+mf}, let us make use of the following identity, which can be derived using the definition of the multivariate Hermite polynomials \eqref{eq:multHermite} and the fact that $V^{(n)}_{k_1k_2\ldots}$ are symmetric in $k_1,k_2,\ldots$,
\begin{equation}\label{eq:hermitesplit}
V^{(n)}_{k_1\ldots k_n}\text{He}_n^{k_1\ldots k_n}(\eta_k) = \sum_{l=0}^n  V^{(n)}_{\underbrace{11\ldots}_{n-l}\underbrace{k_1k_2\ldots}_{l}} \binom{n}{l}\text{He}_{n-l}(\eta_1)\text{He}_l(\eta_k^*),
\end{equation}
where $\binom{n}{l}$ are binomial coefficients and $\eta_k^*$ is a set of all common factors but $\eta_1$.
Using the above expression, the portfolio value \eqref{eq:portmult} can be written as
\begin{equation}\label{eq:tailhermite}
V(\eta_k) = \sum_n\sum_{k_1,k_2,\ldots}\sum_{m\geq n}\binom{m}{n}\text{He}_{m-n}(\eta_1)V^{(m)}_{\underbrace{11\ldots}_{m-n}\underbrace{k_1k_2\ldots}_n}
\text{He}_{n}^{\overbrace{k_1k_2\ldots}^n}(\eta_k^*).
\end{equation}
Finally, separating the $n=0$ term and introducing conditional coefficients ${V^{(n)}_{mf}}_{k_1k_2\ldots}(\eta_1)$, the portfolio value can be put into the form
\begin{equation}\label{eq:porttail}
V(\eta_k)  =  V_{1f}(\eta_1) + V_{mf}(\eta_k^*|\eta_1)
\end{equation}
\vspace{0.1cm}
\begin{equation}\label{eq:porttail2}
V_{1f}(\eta_1)  =  \sum_nV^{(n)}_{1f}\text{He}_n(\eta_1), \quad V_{mf}(\eta_k^*|\eta_1)  =  \sum_{n>0}\sum_{k_1,k_2,\ldots} {V^{(n)}_{mf}}_{k_1k_2\ldots}(\eta_1)\text{He}_n^{k_1k_2\ldots}(\eta_k^*)
\end{equation}
\begin{equation}\label{eq:porttail3}
V^{(n)}_{1f}=V^{(n)}_{\underbrace{11\ldots}_n}, \quad {V^{(n)}_{mf}}_{k_1k_2\ldots}(\eta_1) = \sum_{m\geq n} \binom{m}{n}\text{He}_{m-n}(\eta_1)V^{(m)}_{\underbrace{11\ldots}_{m-n}\underbrace{k_1k_2\ldots}_n}
\end{equation}
The multi-factor correction $V_{mf}$ in the above has zero expectation conditional on $\eta_1$ due to the orthogonality properties \eqref{eq:hermiteorth}. For a given confidence level $\alpha$, the above expressions represent series expansion of the conditional (on $\eta_1=\Phi^{-1}(\alpha)$) tail expectation.

\subsection{Systematic tail risk and its allocation}\label{sec:sysalloc}
The series expansion of the conditional tail expectation \eqref{eq:porttail}-\eqref{eq:porttail3} together with the single-factor case results \eqref{eq:1fvares} allow us to apply the results of Section \ref{sec:adjustments} to VaR and ES calculations.

Since the single-factor VaR and ES have been calculated before, i.e. \eqref{eq:1fvares}, let us start with the second order contributions \eqref{eq:VaR2} and \eqref{eq:ES2}. Using the notations introduced in the previous section, the second order VaR and ES adjustments are
\begin{eqnarray}\label{eq:VaRES2}
\Delta\text{VaR}_2(\alpha) & = & \frac{1}{2V_{1f}'(\eta_1)}\left( \mu_2'(\eta_1) - \mu_2(\eta_1)\left(\eta_1 + \frac{V_{1f}''(\eta_1)}{V_{1f}'(\eta_1)}\right)\right)\Bigg|_{\eta_1=\Phi^{-1}(\alpha)} \\
\Delta\text{ES}_2(\alpha) & = & \frac{1}{2\alpha}\frac{n(\eta_1)}{V_{1f}'(\eta_1)}\mu_2(\eta_1)\Big|_{\eta_1=\Phi^{-1}(\alpha)}\nonumber
\end{eqnarray}
The $V_{1f}$ derivatives can be calculated using \eqref{eq:porttail2} and are
\begin{equation}
V_{1f}'(\eta_1) = \sum_{n>0}V^{(n)}_{1f}n\text{He}_{n-1}(\eta_1), \quad V_{1f}''(\eta_1) = \sum_{n>1}V^{(n)}_{1f}n(n-1)\text{He}_{n-2}(\eta_1)
\end{equation}
The conditional second central moment (variance) $\mu_2(\eta_1)$ is
\begin{equation}\label{eq:mu2}
\mu_2(\eta_1) = \sum_{n>0}n!\sum_{k_1,k_2,\ldots}\Big[{V^{(n)}_{mf}}_{k_1k_2\ldots}(\eta_1)\Big]^2
\end{equation}
and its derivative $\mu_2'(\eta_1)$ can be calculated as
\begin{equation}
\mu_2'(\eta_1) = 2\sum_{n>0}n!\sum_{k_1,k_2,\ldots}{V^{(n)}_{mf}}_{k_1k_2\ldots}(\eta_1) \Big[{V^{(n)}_{mf}}_{k_1k_2\ldots}(\eta_1)\Big]',
\end{equation}
where
\begin{equation}
\Big[{V^{(n)}_{mf}}_{k_1k_2\ldots}(\eta_1)\Big]' = \sum_{m > n} \binom{m}{n}(m-n)\text{He}_{m-n-1}(\eta_1)V^{(m)}_{11\ldots k_1k_2\ldots}
\end{equation}

The above solves the problem of second order VaR and ES adjustments on portfolio level. The corresponding risk contributions can be calculated by applying \eqref{eq:Euler} to \eqref{eq:VaRES2}. This exercise is left for the reader who may find useful the following examples
\begin{eqnarray}\label{eq:eulerderiv}
w_i\frac{\partial}{\partial w_i} V_{1f}'(\eta_1) & = & \sum_{n>0}\frac{\rho_i^n}{n!}v_i^{(n)}(\beta_i)_1^nn\text{He}_{n-1}(\eta_1)\\
w_i\frac{\partial}{\partial w_i} \mu_2(\eta_1) & = & \!\!\!\! 2\sum_{n>0}\rho_i^n\sum_{k_1,k_2,\ldots}{V^{(n)}_{mf}}_{k_1k_2\ldots}(\eta_1) \sum_{m\geq n} \binom{m}{n}\text{He}_{m-n}(\eta_1)v_i^{(m)}(\beta_i)_1^{m-n}(\beta_i)_{k_1}(\beta_i)_{k_2}\ldots \nonumber
\end{eqnarray}

Calculations of the third order VaR and ES adjustments, \eqref{eq:VaR3} and \eqref{eq:ES3}, and corresponding risk contributions can be done in the same fashion. The difficulty one will face in this case is calculation of $\mu_3(\eta_1)$. To calculate the third central moment the following integral has to be evaluated
\begin{equation}
\mu_3(\eta_1) = \int \Big[V_{mf}(\eta_k^*|\eta_1)\Big]^3 \frac{e^{-\sum_{k=2}^{N_f}\eta_k^2/2}}{(2\pi)^{(N_f-1)/2}}\mathrm{d}\eta_k^*
\end{equation}
Unlike the case of $\mu_2(\eta_1)$, orthogonality conditions \eqref{eq:hermiteorth} alone are not sufficient to calculate the integral. One is facing the problem of calculating exponentially weighted average of three Hermite polynomials. To solve this problem, let us start with the following identity (which follows from a more general result of \citet{AssociateHermite})
\begin{equation}
\text{He}_n(x)\,\text{He}_m(x) = \sum_k\binom{n}{k}\binom{m}{k}\,k!\,\text{He}_{n+m-2k}(x)
\end{equation}
The integral then can be solved as follows
\begin{equation}
\int\!\text{d}x\,\text{He}_n(x)\,\text{He}_m(x)\,\text{He}_k(x)\,\frac{e^{-x^2/2}}{\sqrt{2\pi}} = \frac{n!\,m!\,k!}{\left(\frac{m+k-n}{2}\right)!\,\left(\frac{k+n-m}{2}\right)!\,\left(\frac{n+m-k}{2}\right)!},
\end{equation}
provided $m+n+k$ is even and each of $m,n,k$ does not exceed the sum and is not less than the absolute value of the other two. Otherwise, the integral is zero.

It is not clear how to write multivariate version of the above identities. However, using the above identities together with the definition of multivariate Hermite polynomials \eqref{eq:multHermite}, one can solve for any given set of $n,m,k$. For example,
\begin{equation}
\int\!\text{He}_1^{k_1}(\eta_k^*)\text{He}_2^{k_2k_3}(\eta_k^*)\text{He}_3^{k_4k_5k_6}(\eta_k^*)  \frac{e^{-\sum_{k=2}^{N_f}\eta_k^2/2}}{(2\pi)^{(N_f-1)/2}}\mathrm{d}\eta_k^* = 6\delta_{k_1k_4}\delta_{k_2k_5}\delta_{k_3k_6}
\end{equation}

First few terms of the third central moment $\mu_3(\eta_1)$ are
\begin{eqnarray}\label{eq:mu3}
\mu_3(\eta_1) & = & 2\sum_{k_1,k_2}{V^{(1)}_{mf}}_{k_1}(\eta_1){V^{(1)}_{mf}}_{k_2}(\eta_1){V^{(2)}_{mf}}_{k_1k_2}(\eta_1) \\
& + & 6\sum_{k_1,k_2,k_3}{V^{(1)}_{mf}}_{k_1}(\eta_1){V^{(2)}_{mf}}_{k_2k_3}(\eta_1){V^{(3)}_{mf}}_{k_1k_2k_3}(\eta_1) \nonumber \\
& + & 8\sum_{k_1,k_2,k_3}{V^{(2)}_{mf}}_{k_1k_2}(\eta_1){V^{(2)}_{mf}}_{k_2k_3}(\eta_1){V^{(2)}_{mf}}_{k_3k_1}(\eta_1) + \ldots \nonumber
\end{eqnarray}

The results presented in this section allow to calculate portfolio-level and facility-level systematic components of VaR and ES. It is easy to see that the necessary amount of calculations is linear in a number of facilities of the portfolio. Moreover, the calculations can easily be parallelized on a multi-processor machines.

\subsection{Numerical results}\label{sec:numerical}
\begin{figure}[h!]
\centering
\ifpdf
$\begin{array}{ccc}
\includegraphics[width=0.4\textwidth,viewport=60 470 440 740,clip]{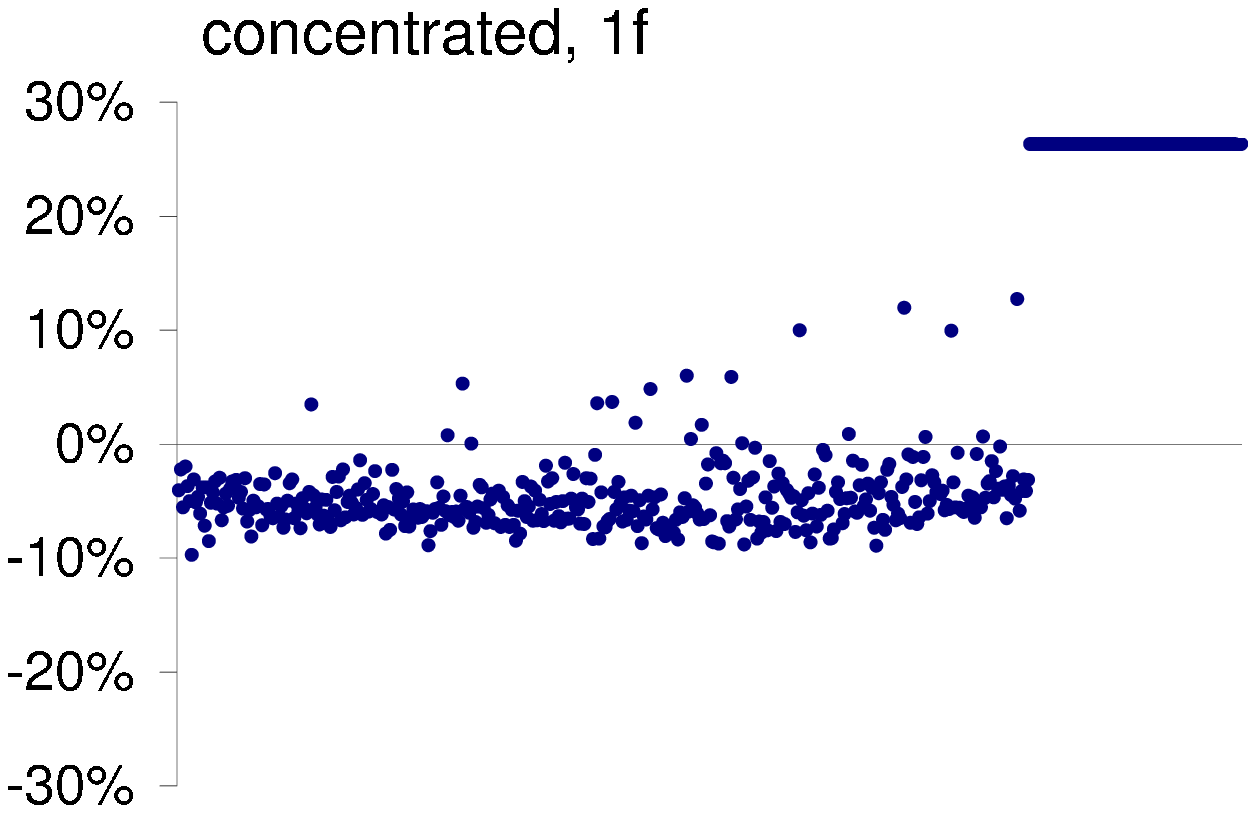} &
\includegraphics[width=0.4\textwidth,viewport=60 470 440 740,clip]{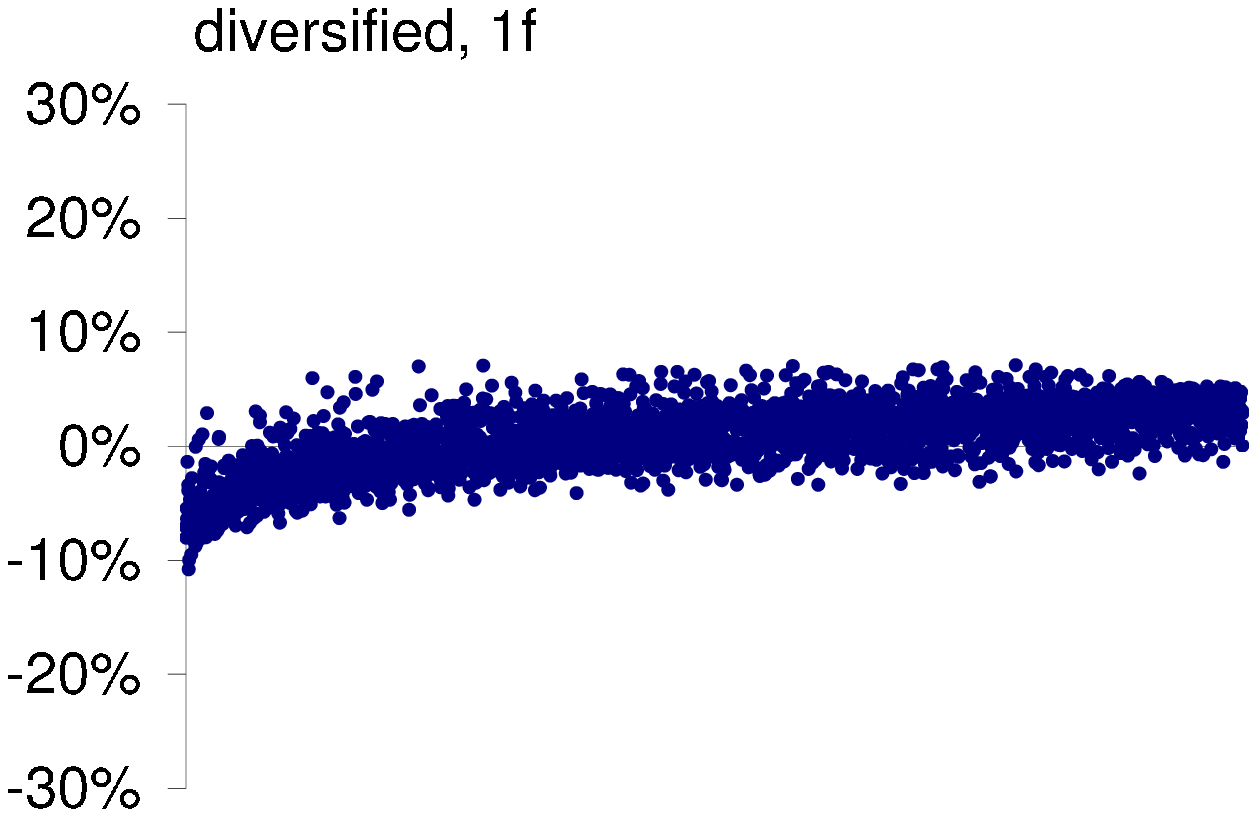} \\
\includegraphics[width=0.4\textwidth,viewport=60 470 440 740,clip]{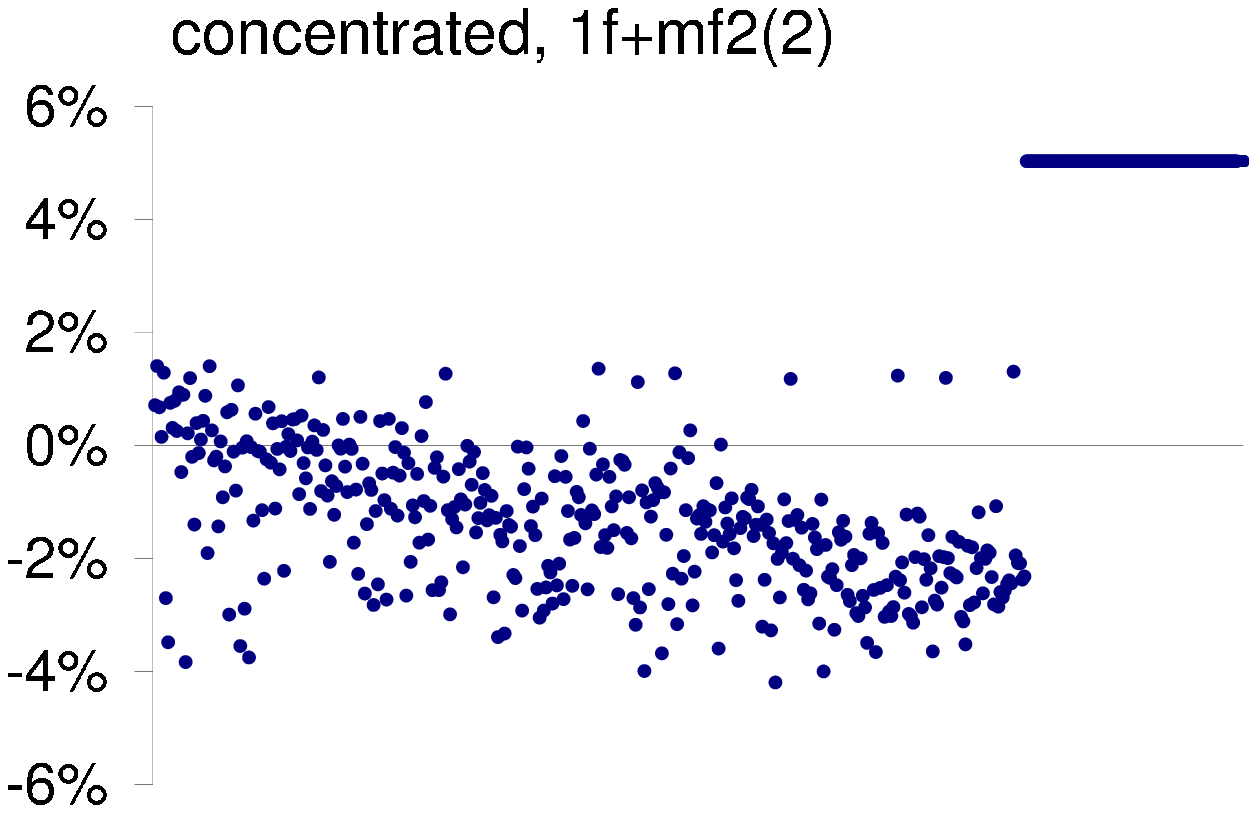} &
\includegraphics[width=0.4\textwidth,viewport=60 470 440 740,clip]{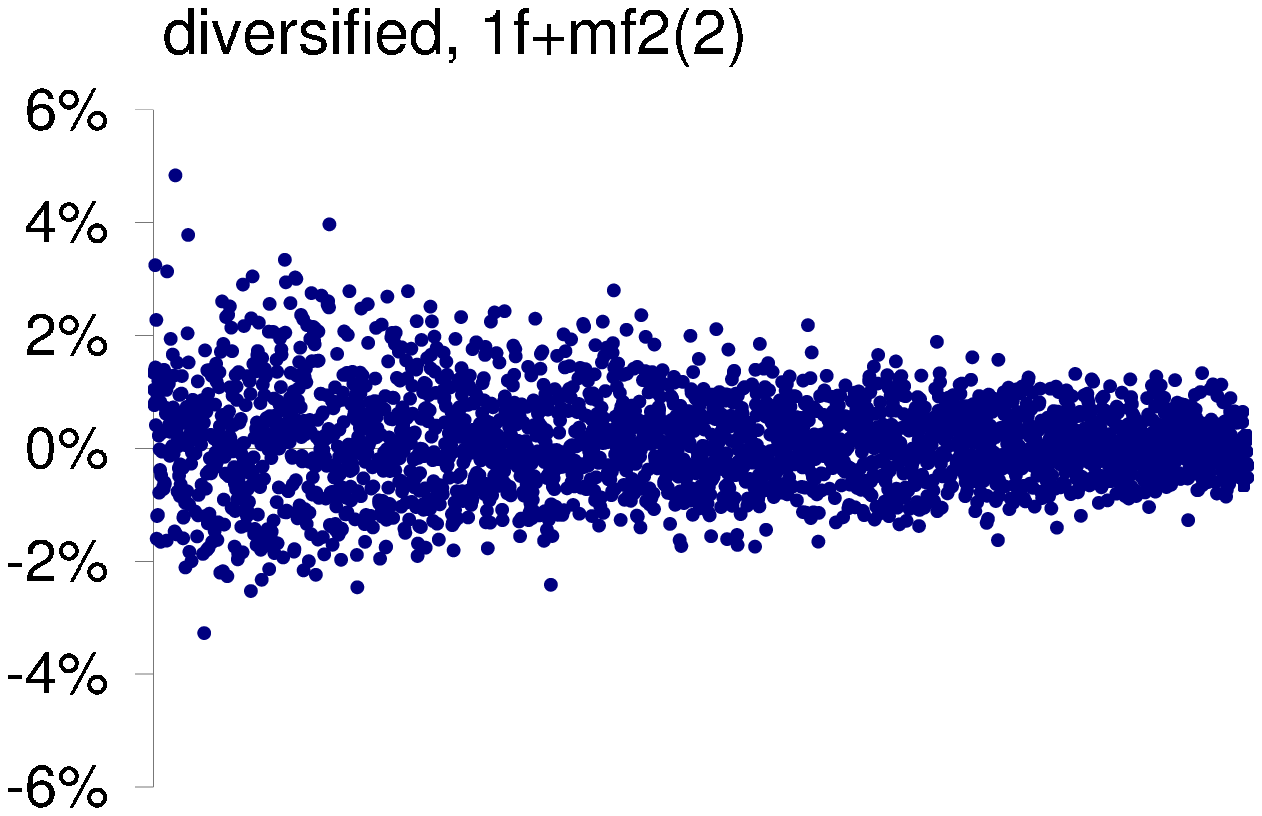} \\
\includegraphics[width=0.4\textwidth,viewport=60 470 440 740,clip]{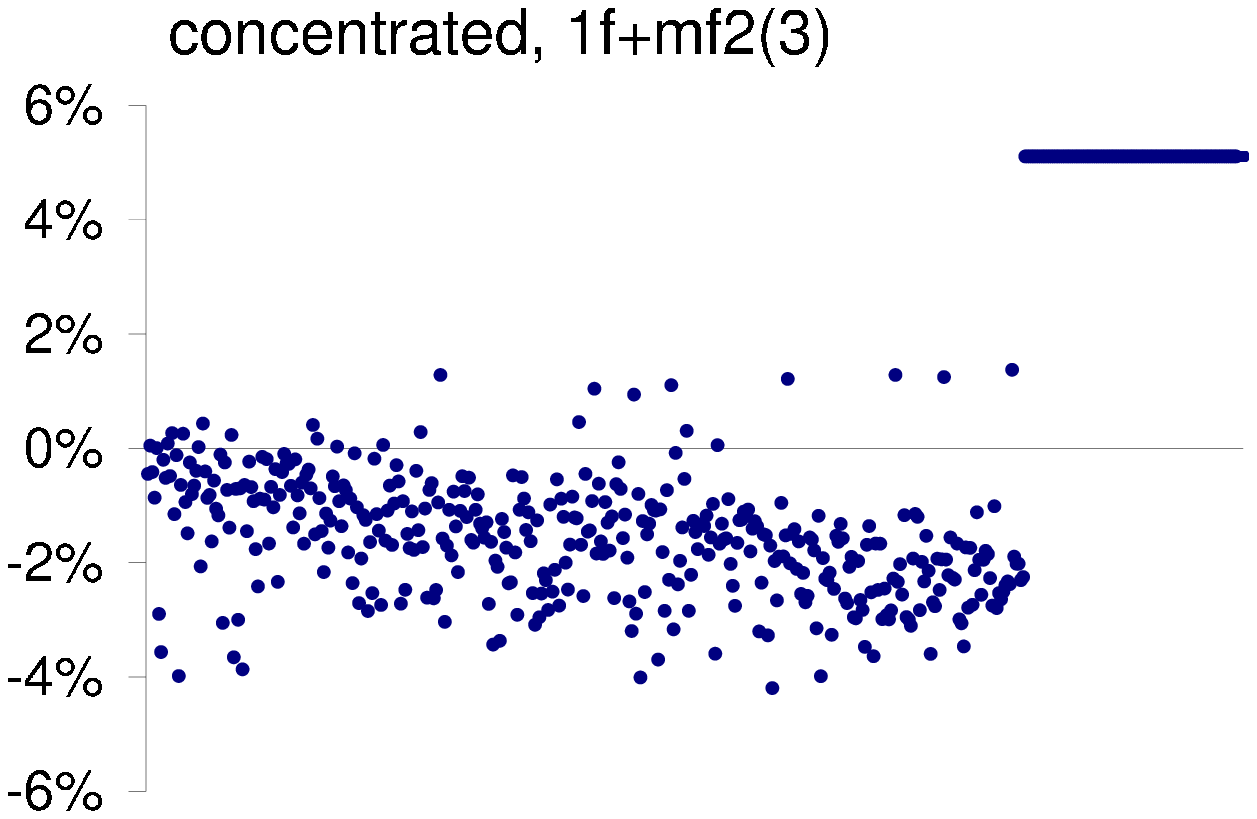} &
\includegraphics[width=0.4\textwidth,viewport=60 470 440 740,clip]{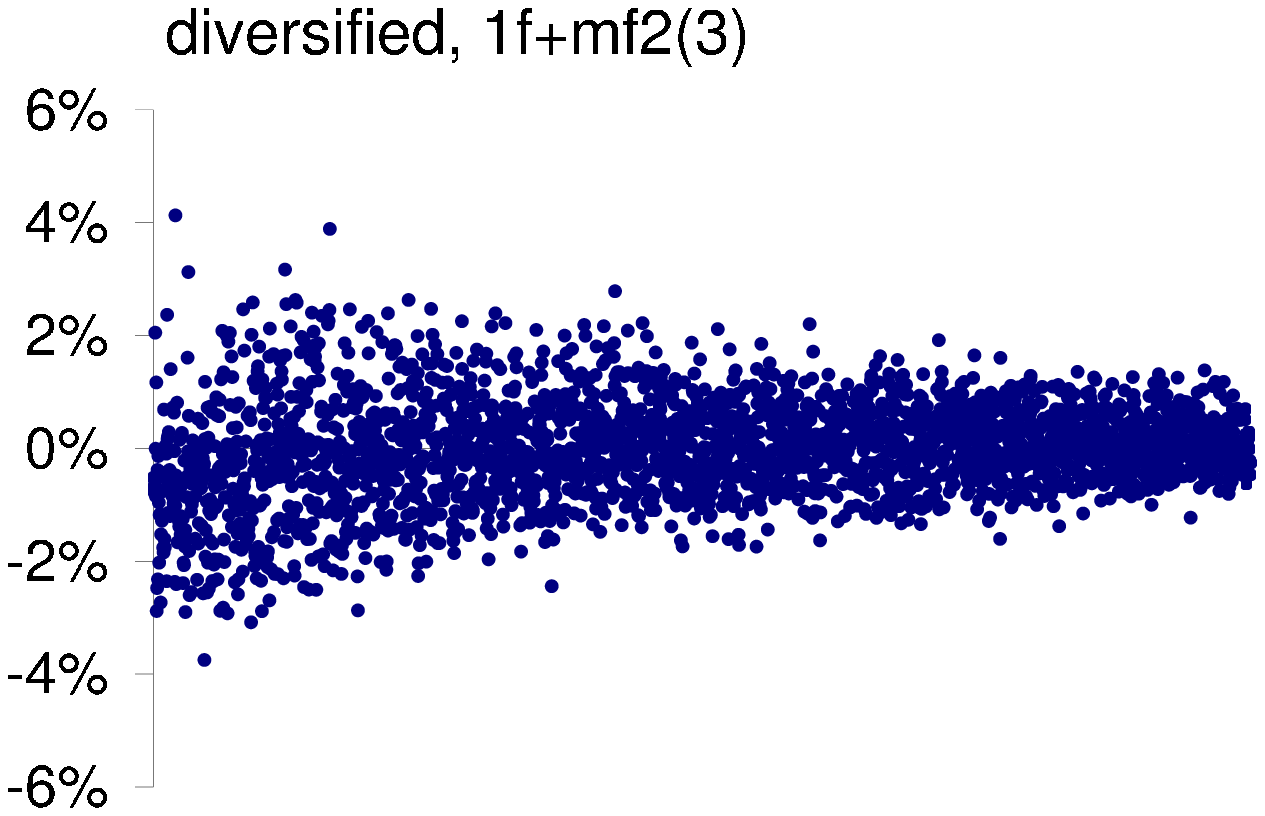} \\
\includegraphics[width=0.4\textwidth,viewport=60 470 440 740,clip]{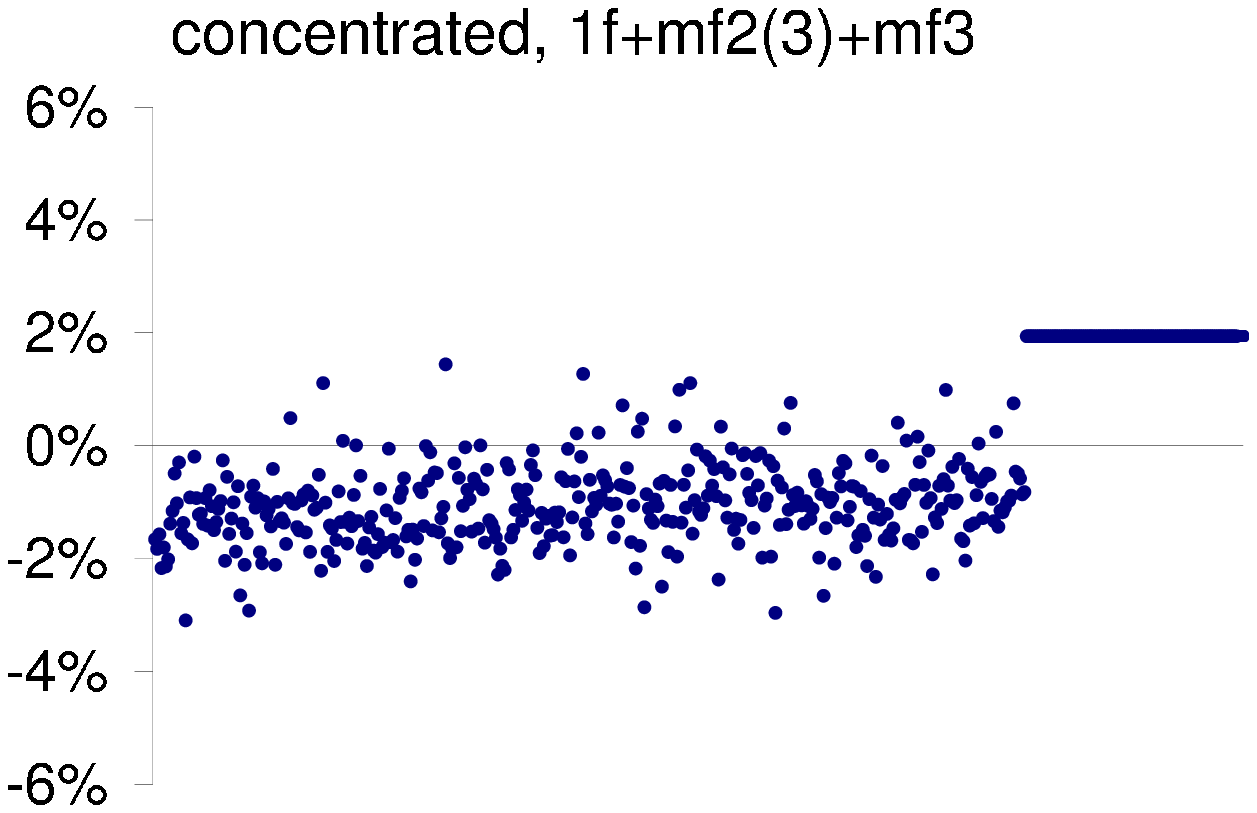} &
\includegraphics[width=0.4\textwidth,viewport=60 470 440 740,clip]{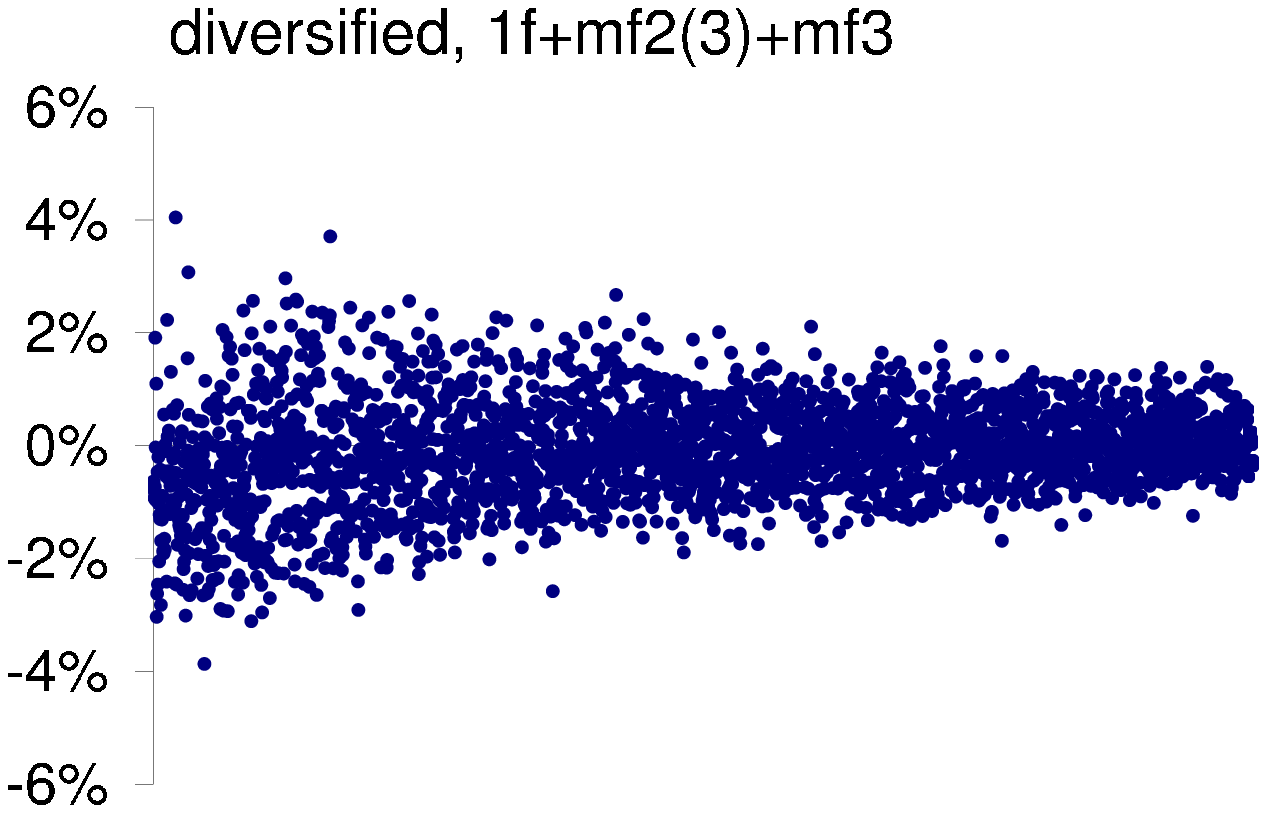}
\end{array}$
\else
$\begin{array}{ccc}
\includegraphics[width=0.4\textwidth,viewport=0 0 380 260,clip]{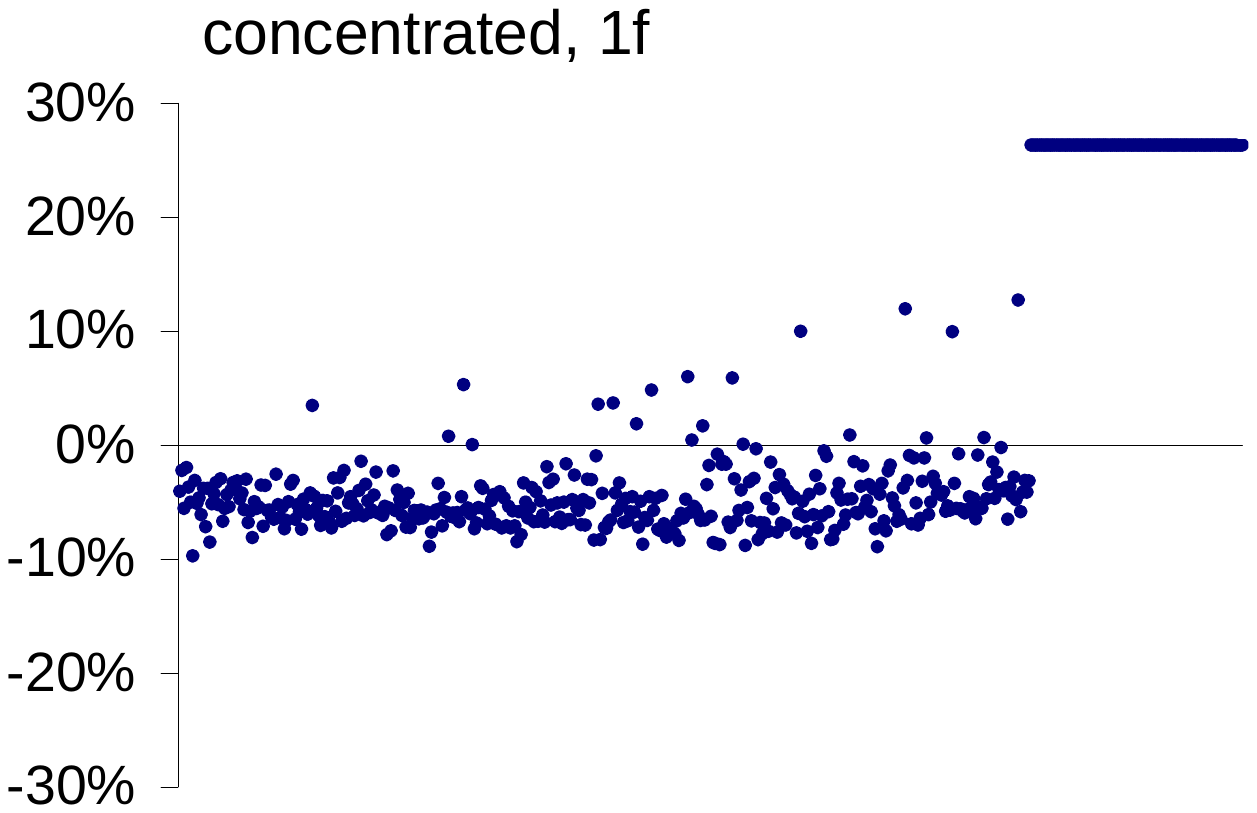} &
\includegraphics[width=0.4\textwidth,viewport=0 0 380 260,clip]{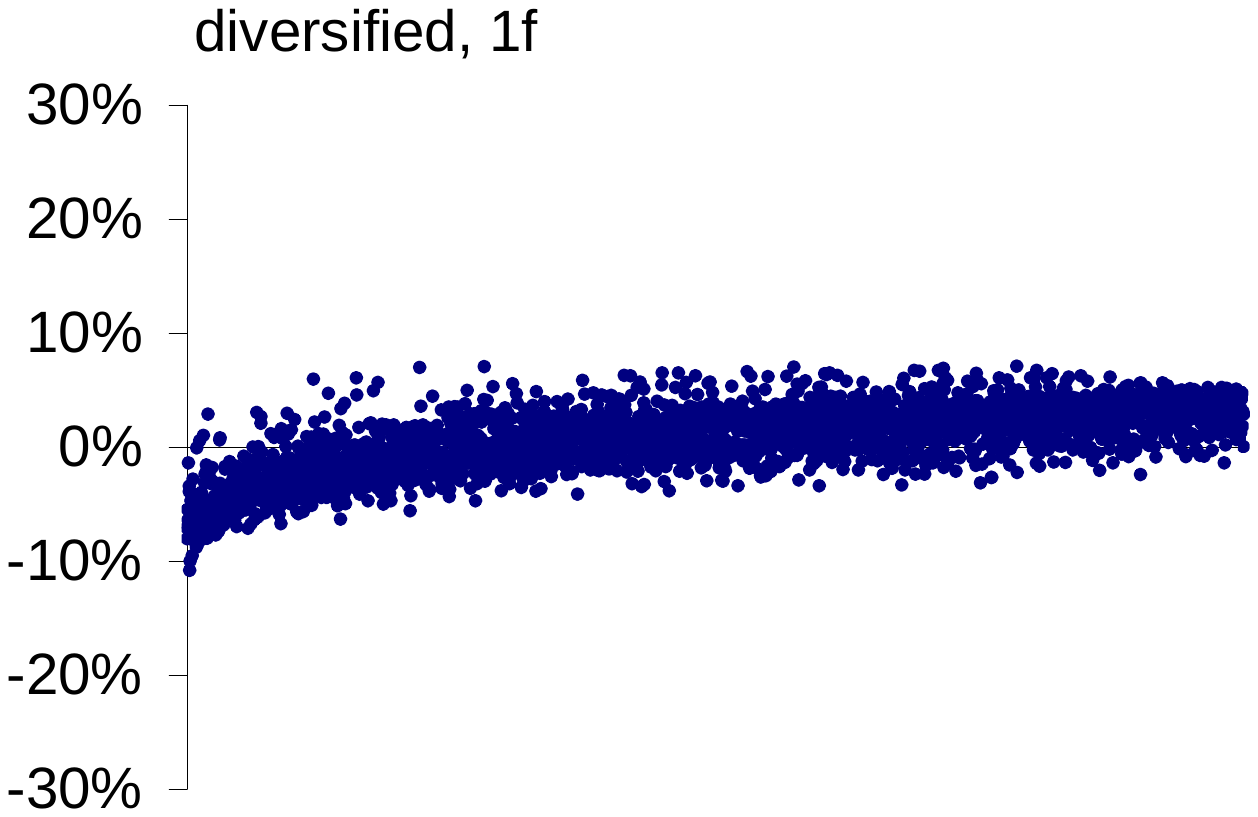} \\
\includegraphics[width=0.4\textwidth,viewport=0 0 380 260,clip]{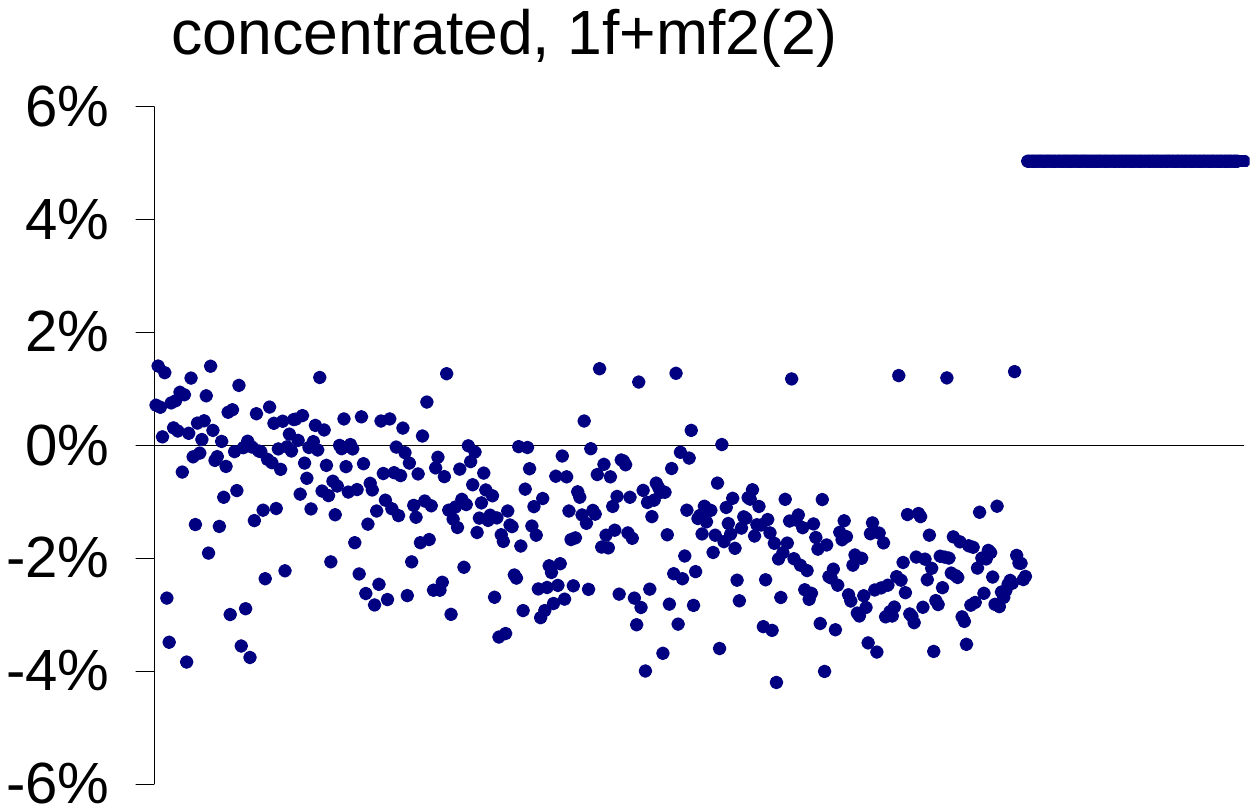} &
\includegraphics[width=0.4\textwidth,viewport=0 0 380 260,clip]{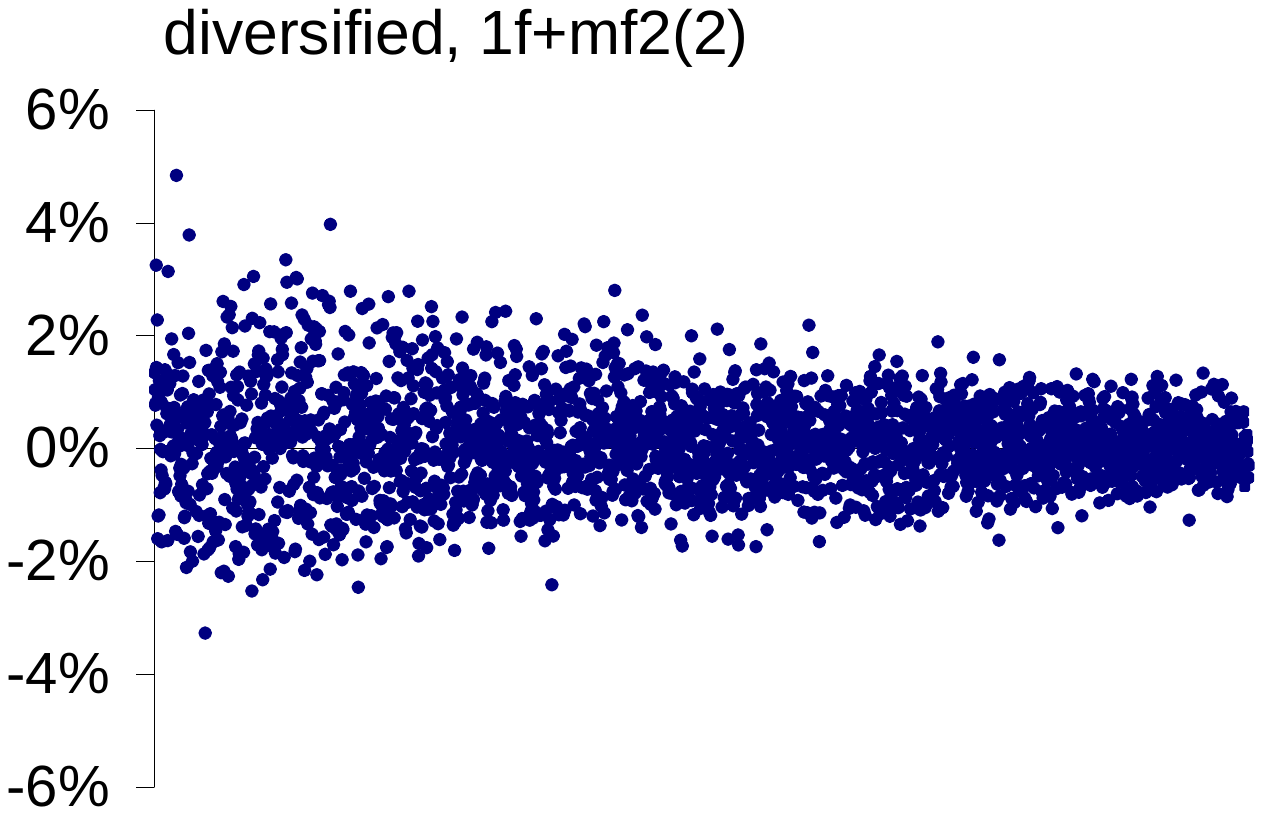} \\
\includegraphics[width=0.4\textwidth,viewport=0 0 380 260,clip]{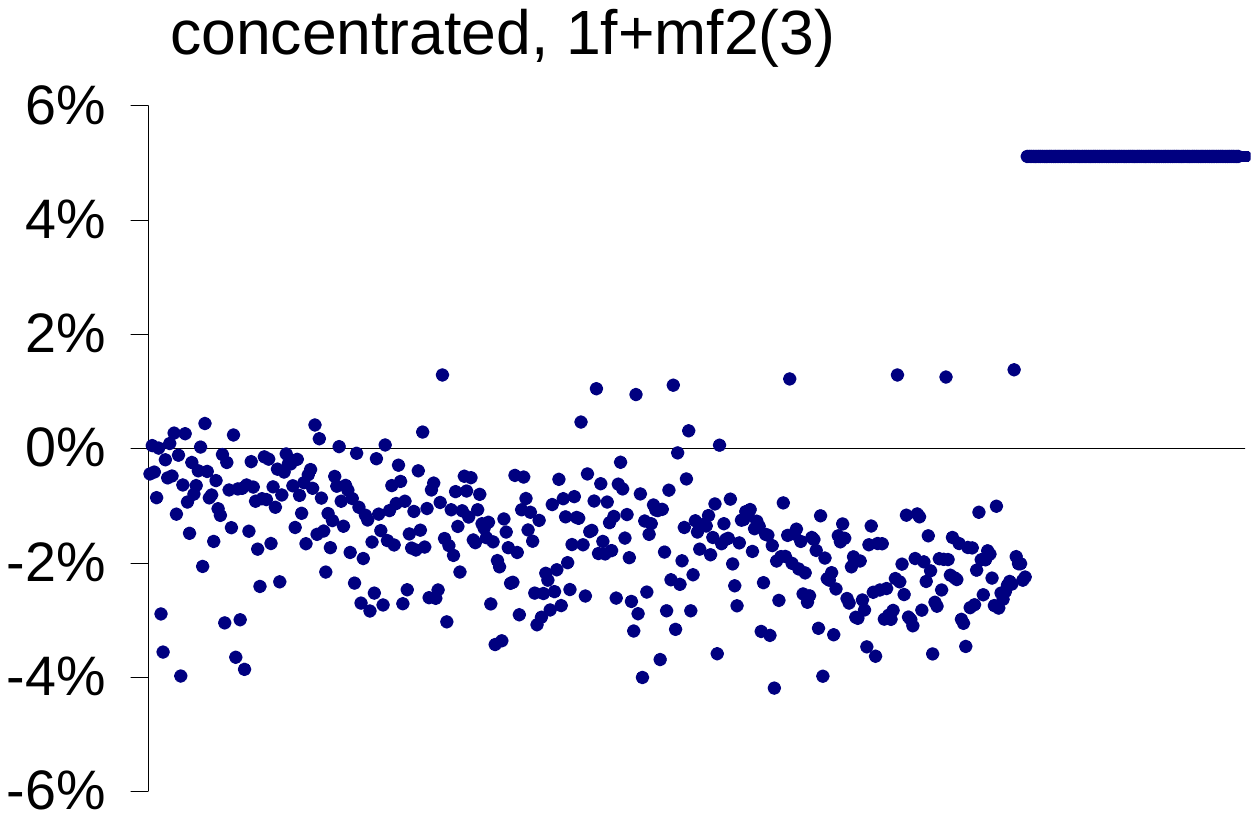} &
\includegraphics[width=0.4\textwidth,viewport=0 0 380 260,clip]{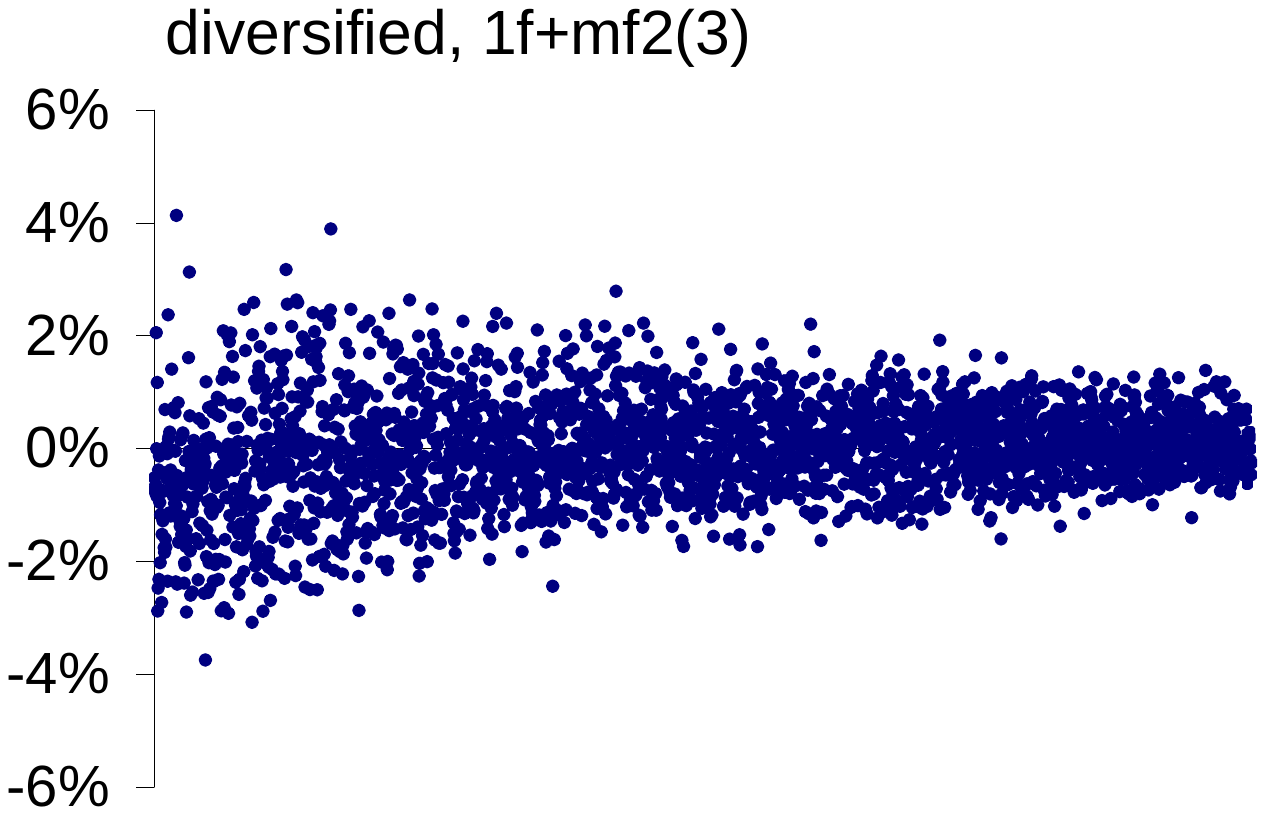} \\
\includegraphics[width=0.4\textwidth,viewport=0 0 380 260,clip]{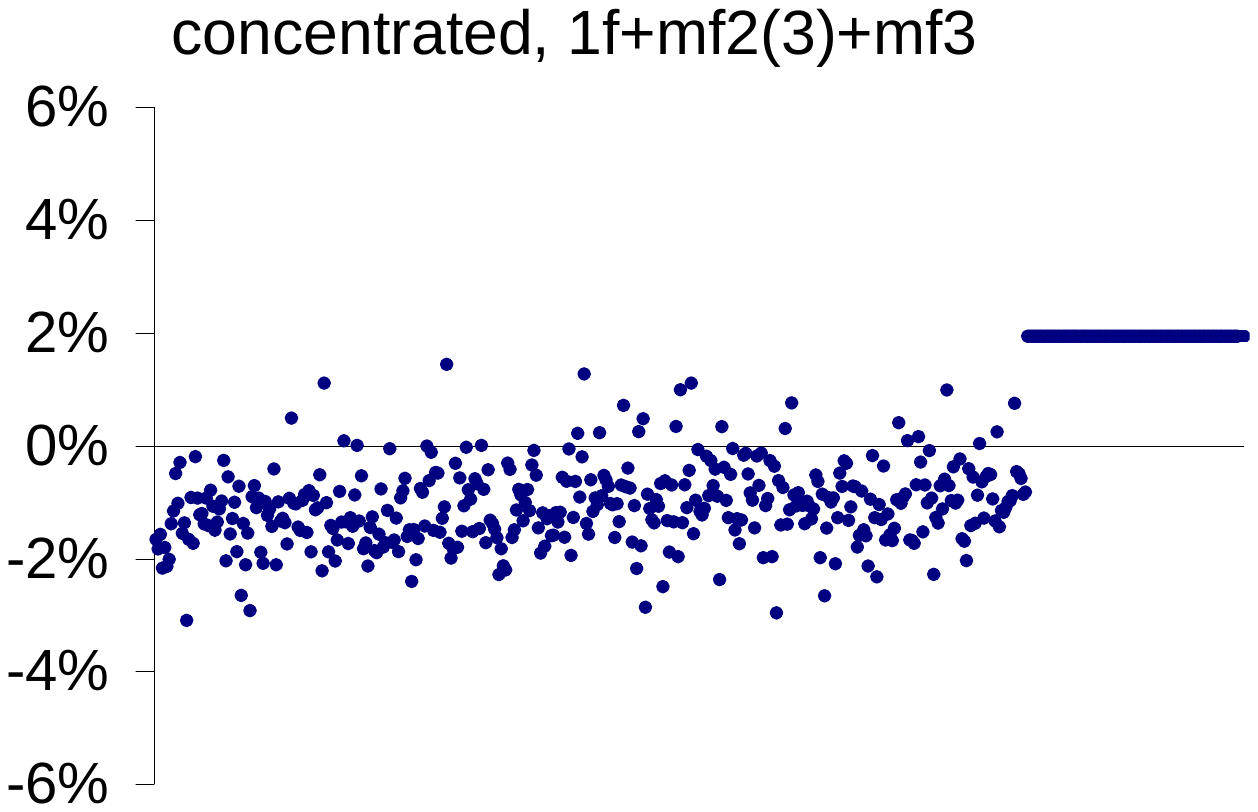} &
\includegraphics[width=0.4\textwidth,viewport=0 0 380 260,clip]{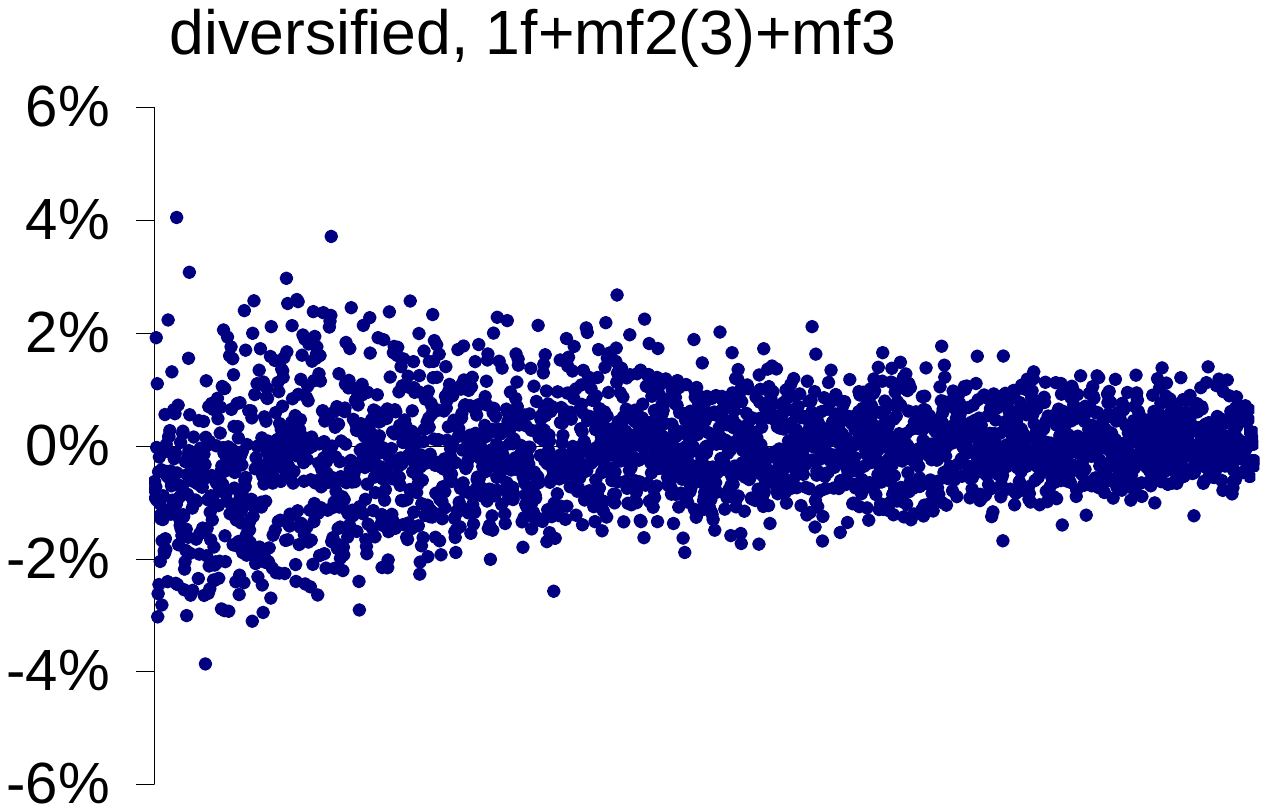}
\end{array}$
\fi
\parbox{0.7\textwidth}{
\caption{\emph{Relative differences between Monte Carlo and analytical estimates of the systematic VaR-based risk contributions.}\label{fig:scatter}}}
\end{figure}

To prove the validity and demonstrate the accuracy of the proposed analytical framework, let us compare results of the analytical approximation with those of unbiased Monte Carlo simulation. The focus here will be on VaR  and VaR-based risk contributions. Only the systematic risk component is considered here. Numerical tests covering idiosyncratic risk as well will be presented in Section \ref{sec:numerical2}.

Since we are interested in the systematic components of portfolio risk, the Monte Carlo routine used here was developed to cover systematic, but not idiosyncratic risk components. This is achieved as follows. For each Monte Carlo scenario a set of systematic factors is generated. Instead of generating borrower-specific factors, however, expected (given systematic factors) values are assigned per facility.

The particular set of common factors used in the tests is similar to the one described in \citet{PortfolioManager}. The total of $N_f=120$ factors cover 61 industry and 45 regional sectors. Two portfolios were constructed, \emph{diversified} and \emph{concentrated}. Both portfolios contain identical loans maturing at horizon. Each loan's correlation with the systematic factors $\rho_i$ is 0.6 and probability of default (PD) equal 1\%. The corresponding value function $v_i(\epsilon)$ is
\begin{equation}
v_i(\epsilon) = \left\{
\begin{array}{rl}
1 & \text{if } \epsilon > \Phi^{-1}(0.01) \\
0 & \text{if } \epsilon \leq \Phi^{-1}(0.01)
\end{array}
\right.
\end{equation}
The \emph{diversified} portfolio contains $45\times61=2745$ loans, each loan representing a different region/industry. The \emph{concentrated} portfolio contains 400 loans randomly assigned to different region/industry and 100 loans representing a single region/industry pair. These 100 loans create region/industry concentration in the portfolio.

Monte Carlo estimates of portfolio VaR and VaR contributions per facility were based on $10^9$ scenarios. Confidence interval was set to 99.9\%. Estimates of VaR contributions were calculated based on 50,000 scenarios around the 99.9\% point (i.e. average VaR contributions for 99.875\% - 99.925\% interval were calculated). Plain vanilla Monte Carlo simulations were used to exclude any bias and limit possibilities of implementation errors.

Several analytical estimates were calculated. First, single factor approximation (\emph{1f}) was calculated based on \eqref{eq:porttail}-\eqref{eq:porttail3} and \eqref{eq:1fvares}-\eqref{eq:1fvaresalloc}. Next, second order (multi-factor) VaR adjustment \eqref{eq:VaR2} was added. The second central moment $\mu_2$ used for calculations was computed using first two (\emph{1f+mf2(2)}) and three (\emph{1f+mf2(3)}) terms in its series expansion \eqref{eq:mu2}.
Finally, analytical estimates were completed by the third order (\emph{1f+mf2(3)+mf3}) VaR adjustment \eqref{eq:VaR3}. The estimation of the third central moment $\mu_3$ was based on the first three terms of its series expansion listed in \eqref{eq:mu3}.

Comparison of the portfolio level results is presented in Table \ref{tab:portfolio}, while VaR-based risk contributions on facility level are compared in Figure \ref{fig:scatter}. The results presented in Figure \ref{fig:scatter} are sorted in increasing from left to right order using a scalar product of the \emph{principal vector} $\vec{Y}$ and factor loadings vectors $\vec{\beta}$ as a parameter.

\begin{table}[h]
\centering
\begin{tabular}{l c c c c}
\hline\hline
& 1f & 1f+mf2(2) & 1f+mf2(3) & 1f+mf2(3)+mf3 \\ \hline \\
\emph{concentrated} & -5.2\% & -0.9\% & -0.8\% & -0.1\% \\
\emph{diversified} & -1.5\% & -0.1\% & -0.1\% & 0.0\% \\ \\ \hline
\end{tabular}
\parbox{0.7\textwidth}{
\caption{\emph{Relative differences between analytical approximation and Monte Carlo simulation on portfolio level.}}
\label{tab:portfolio}}
\end{table}

The following conclusions can be drawn based on the results of the numerical tests. Overall, the analytical approximation produces excellent results. On a portfolio level, a slight underestimation of VaR (economic capital) by a single factor approximation is observed for concentrated portfolios. The situation is improved by higher order corrections, whose contributions lead to very precise results.
In case of VaR contributions, the higher order corrections to the single factor approximation (second and third order VaR adjustments) are are necessary to achieve high accuracy. The resulting analytical estimates of the VaR contributions are just 1-2\% different from the Monte Carlo based estimates.

\section{Idiosyncratic risk}\label{sec:idiosyncratic}
The asymptotic multi-factor framework described in the previous sections was built on the conditional expectation series expansion. As a result, the idiosyncratic risk of the portfolio has been wiped out (averaged over) and portfolio risk measures were expressed in terms of the systematic components. In this section it is shown how the framework described so far can be extended to cover the idiosyncratic risk components.

\subsection{Idiosyncratic contributions}
Using the notations of the previous section and introducing the idiosyncratic value component $V_{ga}$, the full portfolio value $V$ can be written as
\begin{equation}
V(\eta_1,\eta^*,\xi) = V_{1f}(\eta_1) + V_{mf}(\eta^*|\eta_1) + V_{ga}(\xi\,|\eta_1,\eta^*),
\end{equation}
where
\begin{subequations}
\begin{eqnarray}
V_{1f}(\eta_1) & = & \left< V(\eta_1,\eta^*,\xi)\right>_{\eta^*,\xi} \\
V_{mf}(\eta^*|\eta_1) & = & \left< V(\eta_1,\eta^*,\xi)\right>_{\xi} -
\left< V(\eta_1,\eta^*,\xi)\right>_{\eta^*,\xi} \\
V_{ga}(\xi\,|\eta_1,\eta^*) & = & V(\eta_1,\eta^*,\xi) - \left< V(\eta_1,\eta^*,\xi)\right>_{\xi}.
\end{eqnarray}
\end{subequations}
The $\left<\ldots\right>$ in the above stands for average. Also, by construction,
\begin{equation}
\left<V_{ga}\right>_{\xi}= 0, \qquad \left<V_{mf}\right>_{\eta^*} = 0.
\end{equation}

Following \citeauthor{MultiFactor}'s \citeyearpar{MultiFactor} approach, one can treat the sum $V_{mf} + V_{ga}$ as a small correction to $V_{1f}$ and apply the results of Section \ref{sec:adjustments}. The VaR and ES contributions are then expressed in terms of $V_{1f}$ and central moments of $V_{mf} + V_{ga}$. For the second and the third central moments considered here, one can write the following
\begin{subequations}
\begin{eqnarray}
\mu_2[V_{mf} + V_{ga}] & = & \mu_2[V_{mf}] + \left< \mu_2[V_{ga}(\xi)] \right>_{\eta^*} \\
\mu_3[V_{mf} + V_{ga}] & = & \mu_3[V_{mf}] + 3\left< V_{mf}\cdot \mu_2[V_{ga}(\xi)] \right>_{\eta^*} + \left< \mu_3[V_{ga}(\xi)] \right>_{\eta^*}. \label{eq:mixedterm}
\end{eqnarray}
\end{subequations}
The first of the above is a well-known low of total variance, while the second one follows from a more general \emph{low of total cumulance} \citep[see, e.g.,][]{cumulance}.

Thus, to compute the idiosyncratic contribution to the second and the third VaR and ES adjustments, one needs to compute the second and the third conditional central moments of the idiosyncratic component $V_{ga}$. Only these moments, not a detailed information abut idiosyncratic components, are needed to complete the task.

\subsection{Conditional idiosyncratic moments}
Taking advantage of the conditional (on systematic factors) independence of the idiosyncratic components, the second and the third idiosyncratic central moments can be computed as a simple sum of individual (from each loan in the portfolio) contributions. The conditional expectation series expansion \eqref{eq:wowmult} can be applied not just to the value , but to its powers:
\begin{subequations}
\begin{eqnarray}
\overline{v_i}(\eta) & = & \sum_n\frac{\rho_i^n}{n!}v_i^{(n)}\overbrace{(\beta_i)_{k_1}(\beta_i)_{k_2}\ldots}^n
\text{He}_n^{\overbrace{k_1k_2\ldots}^n}(\eta_k) \\
\overline{v^2_i}(\eta) & = & \sum_n\frac{\rho_i^n}{n!}w_i^{(n)}\overbrace{(\beta_i)_{k_1}(\beta_i)_{k_2}\ldots}^n
\text{He}_n^{\overbrace{k_1k_2\ldots}^n}(\eta_k) \\
\overline{v^3_i}(\eta) & = & \sum_n\frac{\rho_i^n}{n!}u_i^{(n)}\overbrace{(\beta_i)_{k_1}(\beta_i)_{k_2}\ldots}^n
\text{He}_n^{\overbrace{k_1k_2\ldots}^n}(\eta_k),
\end{eqnarray}
\end{subequations}
where
\begin{eqnarray}
v_i^{(n)} = \int\!v_i(\epsilon)\text{He}_n(\epsilon)\frac{e^{-\epsilon^2/2}}{\sqrt{2\pi}}\text{d}\epsilon, \quad
w_i^{(n)} = \int\!v^2_i(\epsilon)\text{He}_n(\epsilon)\frac{e^{-\epsilon^2/2}}{\sqrt{2\pi}}\text{d}\epsilon, \quad
u_i^{(n)} = \int\!v^3_i(\epsilon)\text{He}_n(\epsilon)\frac{e^{-\epsilon^2/2}}{\sqrt{2\pi}}\text{d}\epsilon
\end{eqnarray}

The above expansions can be used to compute the conditional central moments of the $i$th loan in the portfolio
\begin{subequations}
\begin{eqnarray}
\left< (\mu_2)_i \right>_{\eta^*} & = & \left<\overline{v^2_i}\right>_{\eta^*}  - \left<\overline{v_i}\cdot \overline{v_i}\right>_{\eta^*} \label{eq:mu2} \\
\left< (\mu_3)_i \right>_{\eta^*} & = & \left<\overline{v^3_i}\right>_{\eta^*}  - 3\left<\overline{v_i}\cdot \overline{v^2_i}\right>_{\eta^*} + 2\left<\overline{v_i}\cdot \overline{v_i}\cdot \overline{v_i}\right>_{\eta^*} \label{eq:mu3}
\end{eqnarray}
\end{subequations}
and the mixed term in \eqref{eq:mixedterm}
\begin{equation}
3\left< V_{mf}\cdot (\mu_2)_i \right>_{\eta^*} = 3\left< V_{mf}\cdot \overline{v^2_i} \right>_{\eta^*} - 3\left< V_{mf}\cdot \overline{v_i}\cdot \overline{v_i} \right>_{\eta^*}
\end{equation}
The averages $\left<\ldots\right>_{\eta^*}$ can be calculated using the techniques developed in Section \ref{sec:systematic}. For example,
\begin{subequations}
\begin{eqnarray}
\left<\overline{v^3_i}\right>_{\eta^*} & = & \sum_n\frac{\beta_1^n}{n!}u^{(n)}\text{He}_n(\eta_1) \\
\left<\overline{v_i}\cdot \overline{v^2_i}\right>_{\eta^*} & = & \sum_{k=0}^\infty k!\left(\sum_{n=k}^\infty\binom{n}{k}v^{(n)}\text{He}_{n-k}(\eta_1) (\beta_1)^{n-k}\left(\sqrt{1-\beta_1^2}\right)^k\right)\times \\
& & \times\left(\sum_{n=k}^\infty\binom{n}{k}w^{(n)}\text{He}_{n-k}(\eta_1) (\beta_1)^{n-k}\left(\sqrt{1-\beta_1^2}\right)^k\right)\nonumber \\
\left<\overline{v_i}\cdot\overline{v_i}\cdot \overline{v_i}\right>_{\eta^*} & = & \sum_{k,l,m} \frac{k!\,l!\,m!}{\left(\frac{l+m-k}{2}\right)!\,\left(\frac{m+k-l}{2}\right)!\,\left(\frac{k+l-m}{2}\right)!} h_kh_lh_m , \\
& & h_k = \sum_{n=k}^\infty\binom{n}{k}v^{(n)}\text{He}_{n-k}(\eta_1) (\beta_1)^{n-k}\left(\sqrt{1-\beta_1^2}\right)^k \nonumber
\end{eqnarray}
\end{subequations}

The central moments $\mu_2$ and $\mu_3$ of the portfolio are calculated by summing over the $\left< (\mu_2)_i \right>_{\eta^*}$ and $\left< (\mu_3)_i \right>_{\eta^*}$ in \eqref{eq:mu2}-\eqref{eq:mu3}. The derivatives of these central moments are trivially calculated in a similar fashion using the property $\mathrm{He}_n'(\eta_1) = n\cdot \mathrm{He}_{n-1}(\eta_1)$. Once the central moments and their derivatives have been calculated, the VaR and ES adjustments are obtained using \eqref{eq:VaR2}-\eqref{eq:VaR3} and \eqref{eq:ES2}-\eqref{eq:ES3}.

Due to the conditional independence, the idiosyncratic contribution is a sum of contributions from individual loans; hence, the amount of calculations needed is linear in a number of loans in the portfolio.

\subsection{Idiosyncratic risk allocation}
Using the techniques presented so far, allocation of idiosyncratic risk is a straightforward (although somewhat laborious) task. The Euler principle \ref{eq:Euler} can be applied to the idiosyncratic risk component in a way similar to the one described in Section \ref{sec:sysalloc}.

Without going into the details of the calculations, let us emphasize the following peculiarity which never receive enough attention in the literature before. As a consequence of the conditional independence concept used, the portfolio level idiosyncratic risk contribution is calculated as a sum of the individual contributions from the underlying loans. These individual contributions, however, are not equal to the risk contributions (referred to as \emph{marginal risk contributions}) calculated using the Euler allocation principle. This is due to the fact that the partial derivative in $w_i\frac{\partial}{\partial w_i}$ is applied not only to the central moments (and their derivatives), but also to the $V', V'', \ldots$ terms in \eqref{eq:VaR2}-\eqref{eq:VaR3} and \eqref{eq:ES2}-\eqref{eq:ES3}. For example, a loan with $\rho_i=1$ in \eqref{eq:Merton}, i.e. with the value depending on systematic risk factors only, gives a zero contribution to the portfolio idiosyncratic moments $\mu_2$ and $\mu_3$. Yet, its contribution may be significant due to the effect mentioned above.

Another interesting empirical property of the marginal idiosyncratic contributions is worth being highlighted\footnote{This is not a generic property of any portfolio. However, the behavior of the marginal contributions described here can be observed in most realistic portfolios.}. Although the overall contribution of the idiosyncratic risk component is positive and not significant for moderately concentrated portfolios, the marginal idiosyncratic contributions may be both positive and negative. It is common for most loans in the portfolio to receive small (compared to systematic part) negative corrections, while a few big exposures receive relatively high positive ones. This counterintuitive behavior is demonstrated in the next section.

\subsection{Numerical results}\label{sec:numerical2}

\begin{figure}[h!]
\centering
\ifpdf
$\begin{array}{ccc}
\includegraphics[width=0.6\textwidth,viewport=55 530 500 730,clip]{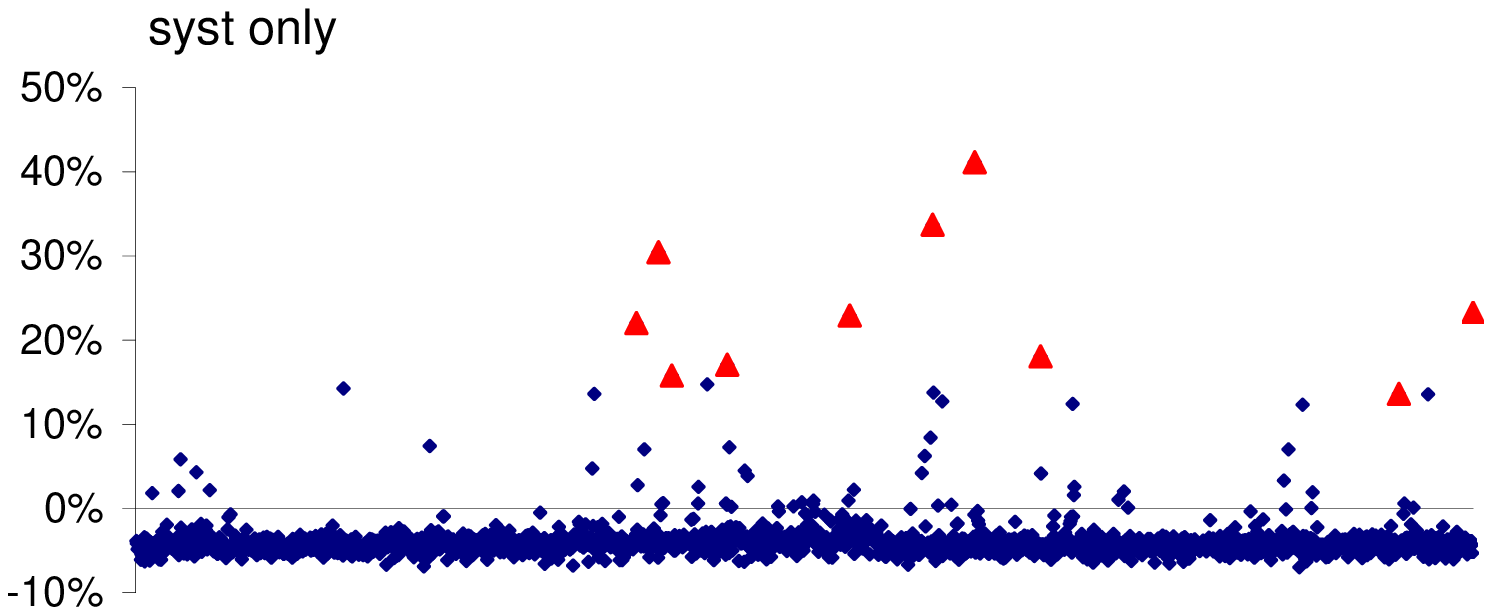} \\
\includegraphics[width=0.6\textwidth,viewport=55 530 500 730,clip]{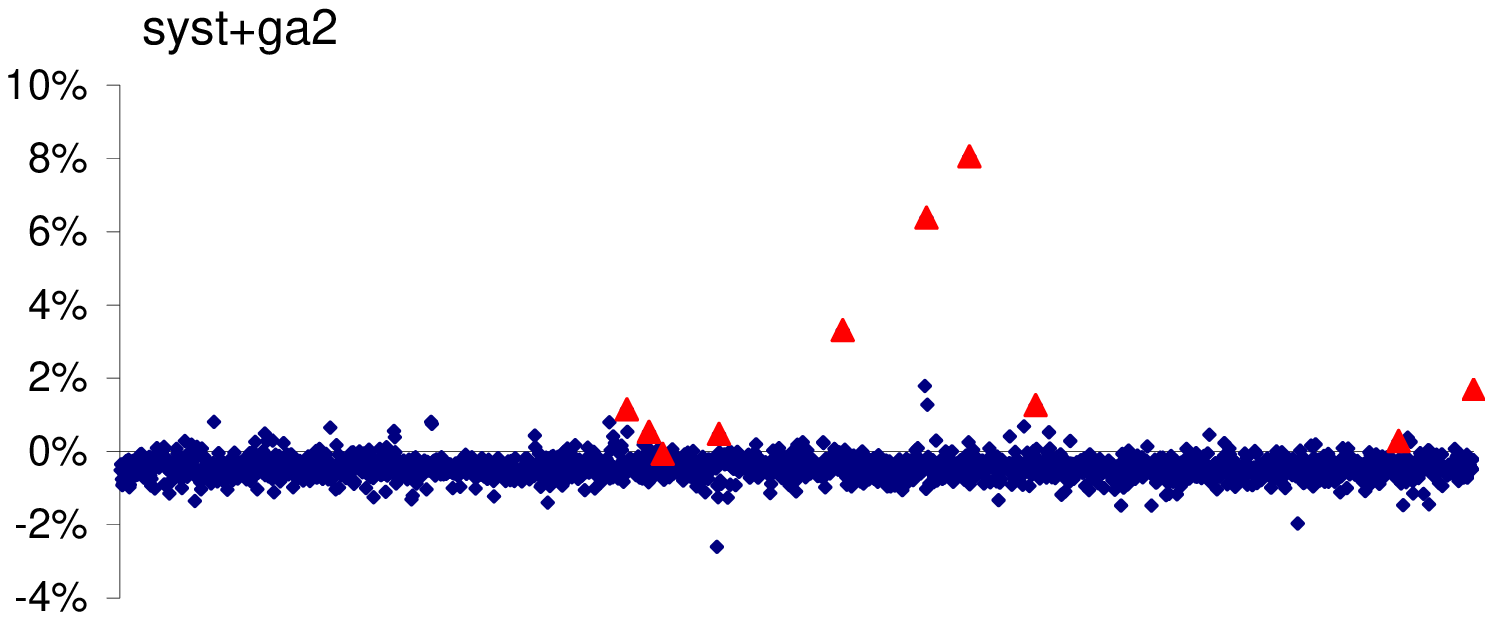} \\
\includegraphics[width=0.6\textwidth,viewport=55 530 500 730,clip]{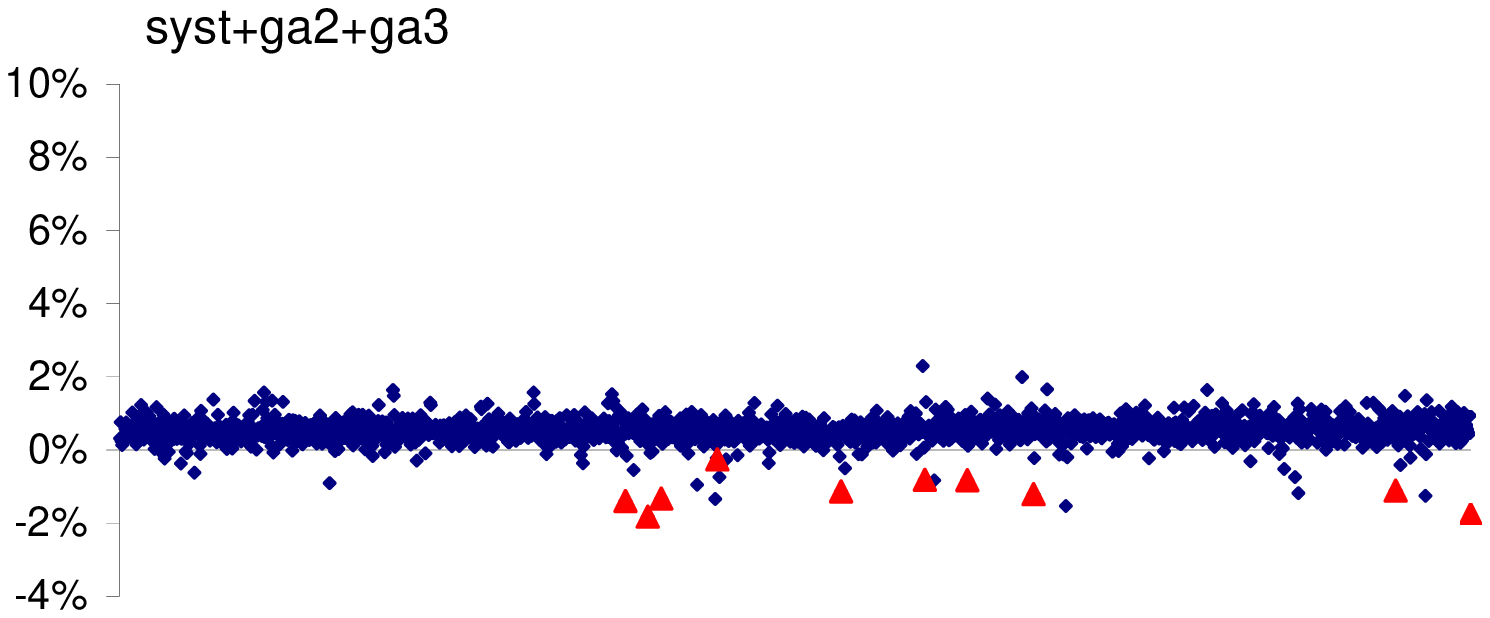}
\end{array}$
\else
$\begin{array}{ccc}
\includegraphics[width=0.6\textwidth,viewport=55 530 500 730,clip]{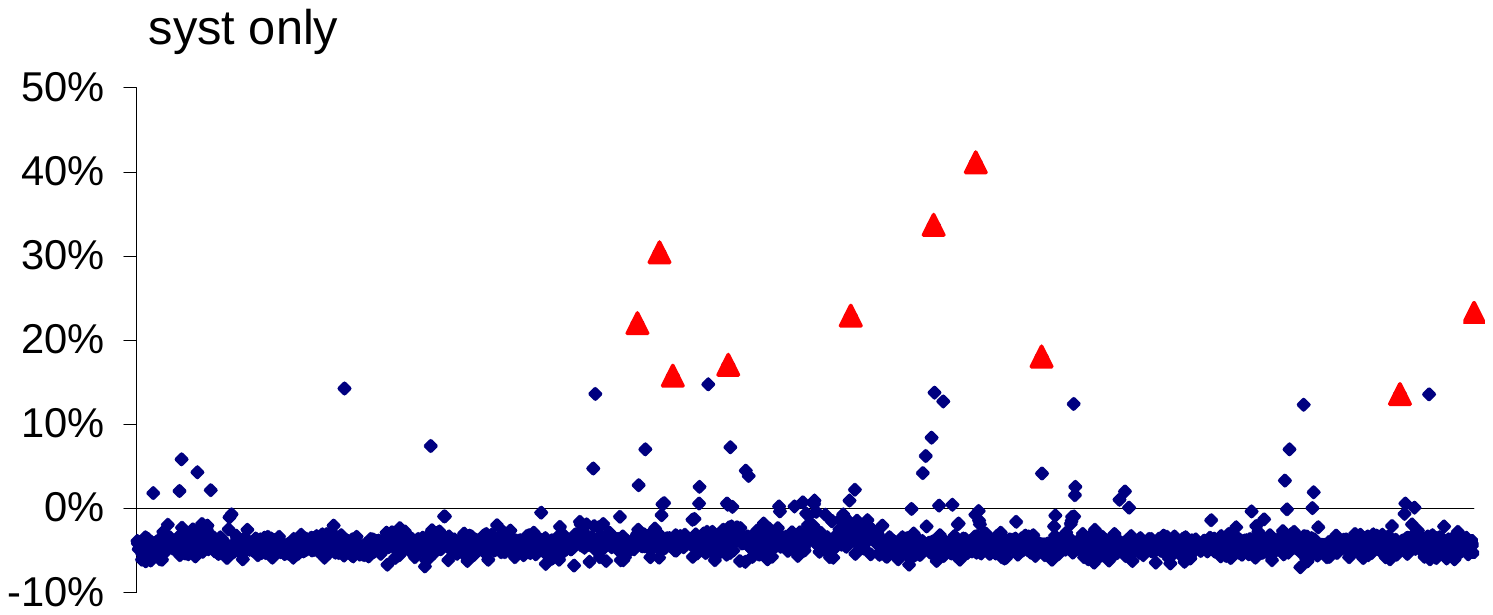} \\
\includegraphics[width=0.6\textwidth,viewport=55 530 500 730,clip]{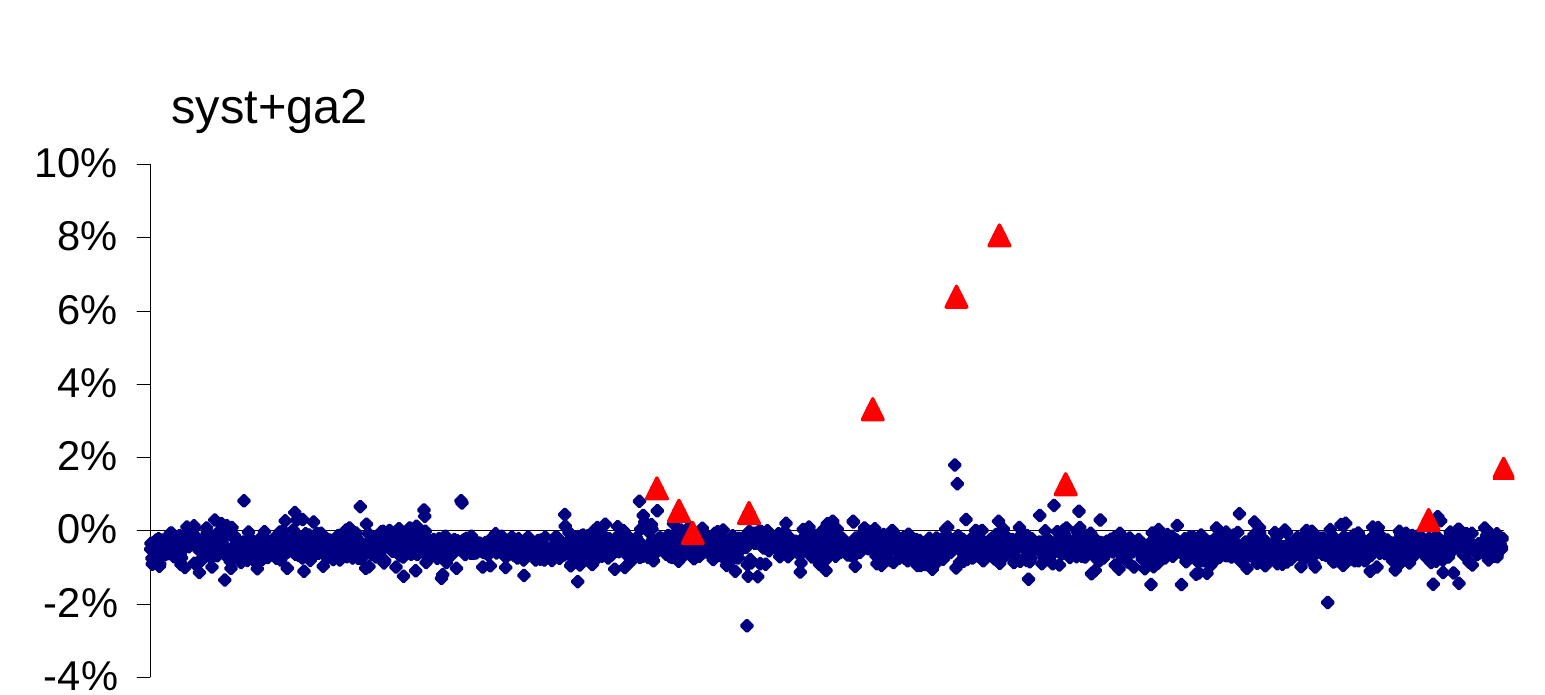} \\
\includegraphics[width=0.6\textwidth,viewport=55 530 500 730,clip]{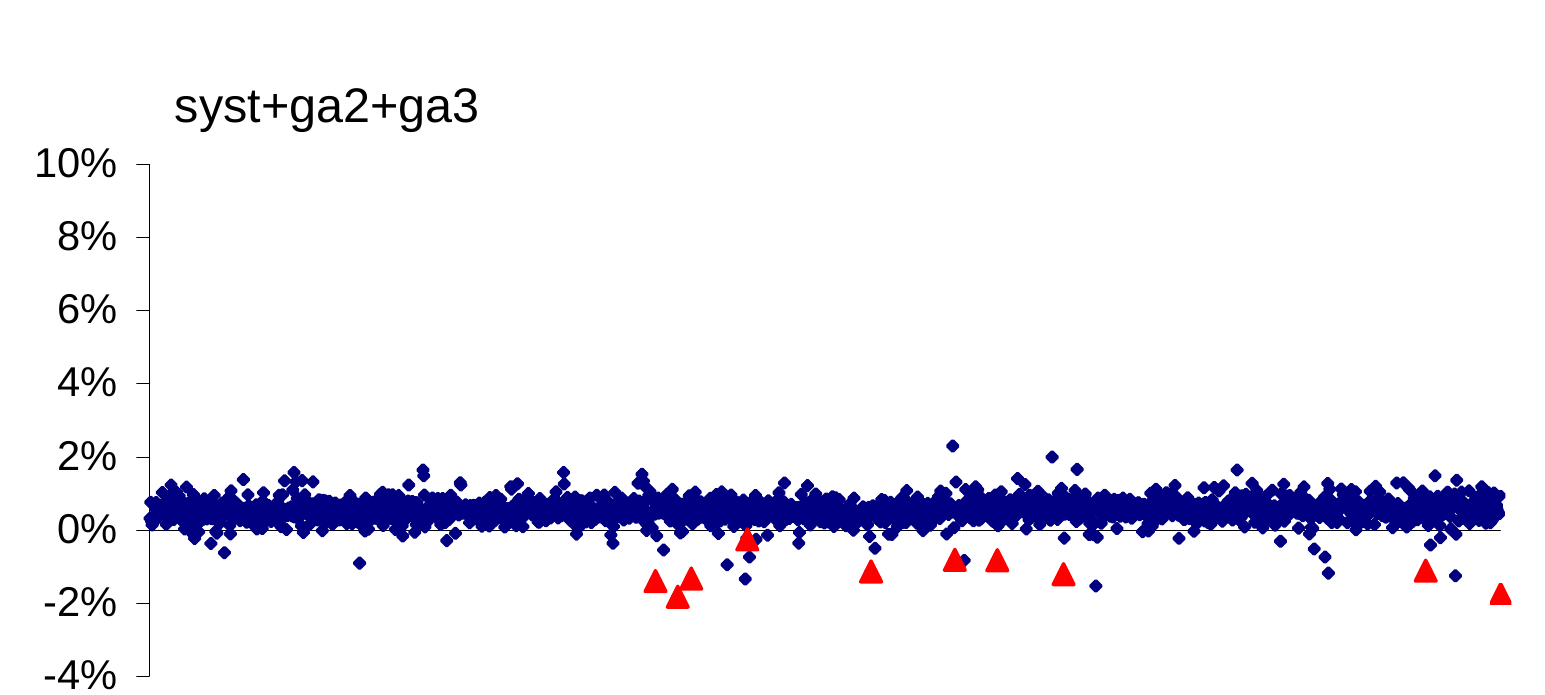}
\end{array}$
\fi
\parbox{0.7\textwidth}{
\caption{\emph{Relative differences between Monte Carlo and analytical estimates of the VaR-based risk contributions.}\label{fig:scatter2}}}
\end{figure}

The analysis presented here mainly focuses on the effects due to the idiosyncratic risk and its concentrations.
In contrast with Section \ref{sec:numerical}, realistic portfolio was used to benchmark the proposed analytical techniques against Monte Carlo simulations. As before, the portfolio VaR and the marginal VaR contributions were the risk measures of interest.

The portfolio consisted of 2,000 loans to distinct customers randomly selected from a loan portfolio of a large European bank. The set of common systematic factors covering 45 geographic regions and 61 regions, as well as the valuation function at horizon $v_i(\epsilon)$ used in the experiment were similar to those of the PortfolioManager \citep[][]{PortfolioManager} model.

Both the portfolio VaR and VaR contributions were estimated using unbiased Monte Carlo simulations. The confidence level was set at 99.9\%. To reduce simulation noise, the VaR contributions were estimated as value expectations in the interval 99.85\%-99.95\%. The simulations consisted of $10^{10}$ (ten billion) scenarios.

The analytical estimates used for comparison with the simulation-based ones were constructed as follows. The systematic part of VaR (\emph{syst}) was used as a starting point. These estimates covered up to the third order VaR adjustment as described in Section \ref{sec:numerical}. The second and the third estimates were computed by consequently adding the second (\emph{ga2}) and the third (\emph{ga3}) order\footnote{The mixed term in \eqref{eq:mixedterm} was included in the third order idiosyncratic contribution. This term gave negligible contribution in the test presented. It is, however, not clear if the term can be neglected in a general case.} idiosyncratic VaR adjustments to the systematic part.

\begin{table}[h!]
\centering
\begin{tabular}{c c c c}
\hline\hline
syst & syst+ga2 & syst+ga2+ga3 \\ \hline \\
-4.1\% & -0.3\% & -0.1\% \\ \hline
\end{tabular}
\parbox{0.7\textwidth}{
\caption{\emph{Relative differences between analytical and simulation-based estimates of the portfolio VaR.}}
\label{tab:portfolio2}}
\end{table}

The relative differences in the portfolio VaRs are presented in Table \ref{tab:portfolio2}. One can observe, that the initial underestimation of VaR by the analytical approximation of systematic risk only is improved significantly by taking into account the second and the third order idiosyncratic contributions. Even without the third order corrections, the portfolio-level results are accurate enough for any practical purposes.

The situation is different if we compare the marginal VaR contributions. Figure \ref{fig:scatter2} shows relative differences in the VaR contributions. The triangles correspond to the ten loans with the highest VaR contributions.
These are the loans with the highest single name risk concentration in the portfolio. The top scatter plot in Figure \ref{fig:scatter2} shows that significant differences between the systematic (analytical) and the full (simulation-based) VaR contributions exist for the concentrated exposures. Both the second and the third idiosyncratic corrections need to be taken into account to obtain precise estimates.

One can also observe the following property of the VaR contributions mentioned before. While a few biggest loans (in terms of VaR contributions) receive significant positive contributions to their VaRs due to idiosyncratic risk, most
loans receive small negative contributions.

Overall, the results are consistent with those of Section \ref{sec:numerical}. Applying the second and the third VaR adjustments to both systematic and idiosyncratic risk components leads to excellent results. On the portfolio level the results may be considered as exact. Computation of the marginal VaR contributions is a more challenging task; however, very high accuracy is achieved by taking into account higher order (the third order in this case) VaR adjustments. 

\section{Summary}\label{sec:summary}
The analytical framework for structured credit portfolio models presented here is an extension and improvement of the one developed by \citet{MultiFactor}. Second and third order VaR and ES adjustments were considered. The default-only case considered by Pykhtin was extended to the case of arbitrary valuation function at horizon. The problem of quadratic (in portfolio size) complexity of Pykhtin's multi-factor adjustment has been solved. High accuracy of the proposed technique was demonstrated by benchmarking with Monte Carlo simulations. The realized performance of the analytical approximation allows it to be considered as not just a supplement, but a substitute to the conventional simulation-based calculations.


\begin{thebibliography}{99}
\bibitem[Abramowitz and Stegun(1972)]{Abramowitz}
M.~Abramowitz and I.~A.~Stegun, eds. \emph{Handbook of Mathematical Functions with Formulas, Graphs, and Mathematical Tables, 9th printing.} New York: Dover, 1972.

\bibitem[Brillinger(1969)]{cumulance}
D.~Brillinger.
\newblock The calculation of cumulants via conditioning.
\newblock \emph{Annals of the Institute of Statistical Mathematics}, Vol. 21, pp.215-218, 1969.

\bibitem[Cespedes et~al.(2006)]{Cespedes}
Cespedes, J.C.G, Herrero, J., Kreinin, A. and D. Rosen (2006)
\newblock A simple multifactor "factor adjustment" for the treatment of credit capital diversification.  \newblock \emph{Journal of Credit Risk}, Vol.2, No.3, Fall 2006.
\newblock Preprint available from
\href{http://www.fields.utoronto.ca/~drosen/Papers/Multi-factor%20Factor%20Adjustment%20-%20January%202006%20FINAL.pdf}{http://www.fields.utoronto.ca/\~{}drosen/Papers/Multi-factor Factor Adjustment - January 2006 FINAL.pdf}

\bibitem[Drake(2009)]{AssociateHermite}
D.~Drake.
\newblock The combinatorics of associate Hermite polynomials.
\newblock \emph{European Journal of Combinatorics}, Vol. 30, No. 4, pp.1005-1021, May 2009.
\newblock Preprint available from
\href{http://www.math.umn.edu/~drake/pdfs/assoc-hermite-fpsac.pdf}{http://www.math.umn.edu/\~{}drake/pdfs/assoc-hermite-fpsac.pdf}

\bibitem[Duellmann and Masschelein(2006)]{Duellmann}
K.~Duellmann and N.~Masschelein.
\newblock Sector concentration risk in loan portfolios and economic capital.
\newblock Working paper, National Bank of Belgium, November 2006.
\newblock Available from \href{http://www.nbb.be/doc/oc/repec/reswpp/WP105.pdf}{http://www.nbb.be/doc/oc/repec/reswpp/WP105.pdf}

\bibitem[Foata(1978)]{Mehler}
D.~Foata.
\newblock A combinatorial proof of the Mehler formula.
\newblock \emph{Journal of Combinatorial Theory}, Series A, Vol. 24, pp. 250-259, 1978.

\bibitem[Gordy(2003)]{ASRF}
M.~Gordy.
\newblock A risk-factor model foundation for ratings-based bank capital rule.
\newblock \emph{Journal of Financial Intermediation}, Vol 12, pp. 199-232, July 2003.
\newblock Preprint available from
\href{http://www.federalreserve.gov/pubs/feds/2002/200255/200255pap.pdf}{http://www.federalreserve.gov/pubs/feds/2002/200255/200255pap.pdf}

\bibitem[Gordy and Marrone(2010)]{GAmtm}
M.~Gordy and J.~Marrone.
\newblock Granularity adjustment for mark-to-market credit risk models.
\newblock Working paper, Federal Reserve Board, June 2010.
\newblock Available from
\href{http://www.federalreserve.gov/pubs/feds/2010/201037/201037pap.pdf}{http://www.federalreserve.gov/pubs/feds/2010/201037/201037pap.pdf}

\bibitem[Gagliardini and Gourieroux(2010)]{GAmf}
P.~Gagliardini and C.~Gourieroux.
\newblock Granularity adjustment in dynamic multiple factor models: systematic vs. unsystematic risk.
\newblock Working paper, March 2010.

\bibitem[Gourieroux et~al.(2000)]{VaRderivatives}
C.~Gourieroux, J.P.~Laurent and O.~Scaillet.
\newblock Sensitivity analysis of values at risk.
\newblock \emph{Journal of Empirical Finance}, Vol. 7, pp 225-245, November 2000.
\newblock Preprint available from
\href{http://sites.uclouvain.be/econ/DP/IRES/2000-2.pdf}{http://sites.uclouvain.be/econ/DP/IRES/2000-2.pdf}

\bibitem[Hull(2007)]{RiskMeasures}
J.~Hull.
\newblock \emph{Risk management and financial institutions.} New Jersey: Pearson Prentice Hall, 2007.

\bibitem[Kalkbrener et~al.(2004)]{Kalkbrener2004}
M.\ Kalkbrener, H.\ Lotter and L.~Overbeck.
\newblock Sensible and efficient capital allocation for credit portfolios.
\newblock \emph{{RISK}}, Vol. 17, pp. 19-24, January 2004.

\bibitem[Kealhofer(2001)]{PortfolioManager}
S.~Kealhofer.
\newblock Portfolio Management of Default Risk.
\newblock Working paper, Moody's KMV, May 2001.
\newblock Available from \href{http://www.moodyskmv.com/research/files/wp/Portfolio_Management_of_Default_Risk.pdf}{http://www.moodyskmv.com/research/files/wp/Portfolio\_Management\_of\_Default\_Risk.pdf}.

\bibitem[Martin and Wilde(2002)]{Unsystematic}
R.~Martin and T.~Wilde.
\newblock Unsystematic credit risk.
\newblock \emph{RISK}, Vol. 15, pp 123-128, November 2002.

\bibitem[Pykhtin(2004)]{MultiFactor}
M.~Pykhtin.
\newblock Multi-factor adjustment.
\newblock \emph{RISK}, Vol. 17, pp. 85-90, March 2004.
\newblock Available from
\href{http://www.riskwhoswho.com/Resources/PykhtinMichael4.pdf}{http://www.riskwhoswho.com/Resources/PykhtinMichael4.pdf}

\bibitem[Tasche(2008)]{Euler}
D.~Tasche.
\newblock Capital allocation to business units and sub-portfolios: the Euler principle.
\newblock Working paper, 2008.
\newblock Available from
\href{http://arxiv.org/PS_cache/arxiv/pdf/0708/0708.2542v3.pdf}{http://arxiv.org/PS\_cache/arxiv/pdf/0708/0708.2542v3.pdf}.

\bibitem[Tasche(2009)]{KernelEstimators}
D.~Tasche.
\newblock Capital allocation for credit portfolios with kernel estimators.
\newblock \emph{Quantitative Finance}, Vol. 9(5), pp. 581-595, August 2009.
\newblock Preprint available from
\href{http://www-m4.ma.tum.de/pers/tasche/Capital_allocation_with_kernel_estimators.pdf}{http://www-m4.ma.tum.de/pers/tasche/Capital\_allocation\_with\_kernel\_estimators.pdf}

\bibitem[Voropaev(2009)]{Voropaev}
M.~Voropaev.
\newblock Variance-covariance based risk allocation in credit portfolios: analytical approximation.
\newblock \emph{{RISK}}, November 2009.
\newblock Preprint available from
\href{http://arxiv.org/abs/0905.0781}{http://arxiv.org/abs/0905.0781}

\end{thebibliography}
\end{document}